\pdfoutput=1
\documentclass[10pt,journal,compsoc]{IEEEtran}
\ifCLASSOPTIONcompsoc
\else
\fi
\ifCLASSINFOpdf
  \usepackage[pdftex]{graphicx}
  \graphicspath{{figures/}}
  \DeclareGraphicsExtensions{.pdf,.jpeg,.png}
\else
\fi
\usepackage[cmex10]{amsmath}
\usepackage{mathtools}
\ifCLASSOPTIONcompsoc
  \usepackage[caption=false,font=normalsize,labelfont=sf,textfont=sf]{subfig}
\else
  \usepackage[caption=false,font=footnotesize]{subfig}
\fi
\usepackage{booktabs}
\usepackage{color}
\usepackage{listings}
\usepackage{url}
\usepackage{xspace}

\newcommand{\code}[1]{\texttt{#1}}

\newcommand{\NumRules}{19\xspace}

\newcommand{\HOU}{\textbf{H}\xspace}
\newcommand{\LU}[1]{\textbf{LU($#1$)}}
\newcommand{\SBASE}{\textbf{SBASE}\xspace}
\newcommand{\SSTEP}{\textbf{SSTEP}\xspace}
\newcommand{\DYN}{\textbf{DYN}\xspace}

\newcommand{\LOOPY}{\textrm{LOOP}\xspace}
\newcommand{\SizeOfLOOPY}{383\xspace}
\newcommand{\SizeOfNonTrivial}{253\xspace}
\newcommand{\SizeOfNonTrivialVerified}{231\xspace}
\newcommand{\SizeOfTrivial}{130\xspace}
\newcommand{\SizeOfTrivialVerifiedWithCandidates}{129\xspace}
\newcommand{\NetGainWithCandidates}{230\xspace}

\newcommand{\NumNotStudied}{330\xspace}
\newcommand{\NumVacuousPass}{82\xspace} 
\newcommand{\NumNotHandled}{10\xspace} 
\newcommand{\NumRequireRefinements}{40\xspace}
\newcommand{\NumNoLoops}{198\xspace}

\begin{document}
%
\title{Implementing and Evaluating Candidate-Based Invariant Generation}
%
%
%
%

\author{Adam~Betts,
            Nathan~Chong,
            Pantazis~Deligiannis,
            Alastair~F.~Donaldson,
            Jeroen Ketema
\IEEEcompsocitemizethanks{
\IEEEcompsocthanksitem A. Betts, N. Chong, P. Deligiannis, A.F. Donaldson and J. Ketema
are with the Department of Computing, Imperial College London, London SW7 2AZ, UK.
\protect\\
E-mail: \{abetts,nyc04,p.deligiannis,afd,j.ketema\}@imperial.ac.uk}
\thanks{}}

%
%

\markboth{}%
{Betts \MakeLowercase{\textit{et al.}}: Scaling Up Candidate-Based Invariant Generation}
%


\IEEEcompsoctitleabstractindextext{%
\begin{abstract}
The discovery of inductive invariants lies at the heart of static program verification.
Presently, many automatic solutions to inductive invariant generation are inflexible, only applicable to certain classes of programs, or unpredictable.
An automatic technique that circumvents these deficiencies to some extent is \emph{candidate-based invariant generation}, whereby a large number of candidate invariants are guessed and then proven to be inductive or rejected using a sound program analyzer.
This paper describes our efforts to apply candidate-based invariant generation in GPUVerify, a static checker for programs that run on GPUs.
We study a set of $\SizeOfLOOPY$ GPU programs that contain loops, drawn from a number of open source suites and vendor SDKs.
Among this set, $\SizeOfNonTrivial$ benchmarks require provision of loop invariants for verification to succeed.

We describe the methodology we used to incrementally improve the
invariant generation capabilities of GPUVerify to handle these
benchmarks, through candidate-based invariant generation,
using cheap static
analysis to speculate potential program invariants.  We also describe
a set of experiments that we used to examine the effectiveness of our
rules for candidate generation, assessing rules based on their
\emph{generality} (the extent to which they generate candidate
invariants), \emph{hit rate} (the extent to which the generated
candidates hold), \emph{worth} (the extent to which provable
candidates actually help in allowing verification to succeed), and
\emph{influence} (the extent to which the success of one generation
rule depends on candidates generated by another rule).  We believe
that our methodology may serve as a useful framework for other researchers interested
in candidate-based invariant generation.

The candidates produced by GPUVerify help to verify $\SizeOfNonTrivialVerified$ of the $\SizeOfNonTrivial$ programs.
This increase in precision, however, makes GPUVerify sluggish: the more candidates that are generated, the more time is spent determining which are inductive invariants.
To speed up this process, we have investigated four under-approximating program analyses that aim to reject false candidates quickly and a framework whereby these analyses can run in sequence or in parallel.
Across two platforms, running Windows and Linux, our results show that the best combination of
these techniques running \emph{sequentially} speeds up invariant
generation across our benchmarks by $1.17\times$ (Windows) and $1.01\times$ (Linux), with per-benchmark
best speedups of $93.58\times$ (Windows) and $48.34\times$ (Linux), and worst slowdowns of $10.24\times$ (Windows) and $43.31\times$ (Linux).  We
find that \emph{parallelizing} the strategies marginally improves overall invariant generation speedups to $1.27\times$ (Windows) and $1.11\times$ (Linux),
maintains good best-case speedups of $91.18\times$ (Windows) and $44.60\times$ (Linux), and, importantly, dramatically reduces worst-case slowdowns to $3.15\times$ (Windows) and $3.17\times$ (Linux).
\end{abstract}

}

\maketitle

\IEEEdisplaynotcompsoctitleabstractindextext

%
\IEEEpeerreviewmaketitle

\section{Introduction}
\label{sec:intro}

An \emph{invariant} is a property that captures program behaviors by expressing a fact that always holds at a particular program point.
Invariants are vital to static verification tools for reasoning about loops and procedure calls in a modular fashion~\cite{floyd1967assigning}.
Such reasoning requires proving that invariants are \emph{inductive}.  In the case of loops this means that they hold on entry to the loop (the base case), and that if they hold at the start of an arbitrary iteration of the loop, they also hold at the end of the iteration (the step case).

The automatic discovery of inductive invariants is a challenging problem that has received a great deal of attention from researchers~\cite{DBLP:conf/cav/GuptaR09,DBLP:conf/fase/KovacsV09,AI77,Houdini01,DBLP:conf/date/Eijk98,DBLP:conf/cav/JeannetM09,AbstractHoudini,DBLP:conf/cav/McMillan06,Abduction13,DBLP:conf/popl/JeannetSS14,DBLP:journals/tse/ErnstCGN01}.
A flexible solution is offered by \emph{candidate-based invariant generation}~\cite{Houdini01,DBLP:journals/ipl/FlanaganJL01} whereby a large number of candidate invariants (henceforth, just \emph{candidates}) are speculated through simple \emph{rules} (e.g.\ based on patterns observed in the abstract syntax tree of a program) and are then checked using formal verification methods.
The output is a subset of the candidates that can be proven to hold; if all candidates are rejected then the weakest invariant, \code{true}, is returned.

Although candidate-based invariant generation is popular in several static verification tools, no general methodology exists to guide the implementation and evaluation of such methods.
We address this problem in this paper by describing the systematic manner in which we incorporated new candidate generation rules into GPUVerify~\cite{GPUVerify12}, a static verification tool for programs that have been designed to run on Graphics Processing Units (GPUs), and by proposing a set of questions that allow rules to be comprehensively evaluated, irrespective of the application domain.
Our evaluation criteria broadly assess whether rules produce inductive invariants, whether there are dependencies among rules, and the extent to which rules help to verify programs.

We applied the proposed evaluation criteria to GPUVerify using a set of $\SizeOfLOOPY$ GPU programs collected from a variety of sources.  This endeavor led to three interesting discoveries.
First, the rules as a whole have greatly increased the accuracy of GPUVerify, helping to verify $\SizeOfNonTrivialVerified$ out of $\SizeOfNonTrivial$ GPU programs where loop invariants are required.
Second, most rules in GPUVerify are independent of each other: if a rule produces inductive invariants, then the proof of those candidates can be established in isolation from inductive invariants of other rules.
Third, some rules in GPUVerify are redundant because they no longer produce candidates that are essential to verify a single GPU program: they have been superseded by more general rules.

Increased precision, however, has come at a price: GPUVerify has become less responsive with the introduction of more rules, because the more candidates that are speculated, the more time is spent determining whether those candidates are actual inductive invariants.
In one specific case, a GPU program that verified within 10 minutes when \emph{no} invariants were speculated, could no longer be verified within 30 minutes due to the overhead of candidate-based invariant generation.
To counter the performance lag, we have investigated four \emph{under-approximating} program analyses whose aim is to refute false candidates quickly, and we have devised a framework where several of these analyses can run either \emph{in sequence} or \emph{in parallel}.
Evaluating these techniques on two different machines, running Windows and Linux, respectively, we discovered that:
\begin{itemize}
  \item In the best case, accelerating invariant generation using a sequential combination of techniques sped up invariant generation performance by $93.58\times$ (Windows) and $48.34\times$ (Linux).

  \item In the worst case, attempts at sequential acceleration did not pay off, slowing down invariant generation by $10.24\times$ (Windows) and $43.31\times$ (Linux).

  \item Over all benchmarks, sequential acceleration sped up invariant generation by $1.17\times$ (Windows) and $1.01\times$ (Linux).

  \item Parallelizing our strategies maintained good best case speedups of $91.18\times$ (Windows) and $44.60\times$ (Linux), while reducing worst-case slow downs to $3.15\times$ (Windows) and $3.17\times$ (Linux).  The key benefit of parallelization here is that it prevents a runaway under-approximating analysis from severely delaying invariant discovery.

\item Overall, parallelization gave a marginally better speedup across our benchmarks, of $1.27\times$ (Windows) and $1.11\times$ (Linux).

\end{itemize}

The rather different results obtained for these distinct platforms emphasize the importance of accounting for measurement bias~\cite{MeasurementBias}: the operating system and programming language runtime used to conduct program analysis can have a noticeable impact on performance results.

In summary, our main contributions are:

\begin{enumerate}

\item An account of a systematic approach to manually deriving domain-specific rules for candidate-based invariant generation.  While the rules are specific to the context of GPU kernel verification, we believe the principled approach we have taken in their discovery can be transferred to other domains.

\item An experimental study of generality, hit-rate, worth, and influence of these candidate generation rules.  We believe that the metrics we present and our experimental method can help to guide other researchers in evaluating candidate-based invariant generation techniques.

\item General strategies for accelerating candidate-based invariant generation via under-approximating program analyses, the application of parallel processing to combine multiple strategies, and a large experimental evaluation of these methods.

\end{enumerate}

The remainder of the paper is structured as follows.
Necessary background material is provided in Section~\ref{sec:background}.
In Section~\ref{sec:benchmarks}, we outline some basic properties of the GPU programs in our evaluation set and describe how preconditions for each GPU program were procured.
The methodology to implement and to evaluate candidate-based invariant generation appears in Section~\ref{sec:rules}, including new metrics for evaluation which we used to assess the candidate generation rules we added to GPUVerify.
The measures undertaken to boost performance and an evaluation thereof are contained in Section~\ref{sec:performance}.
We survey related work in Section~\ref{sec:related} and conclude in Section~\ref{sec:conclusions}.

\section{Background}
\label{sec:background}

We give essential background on loop invariants and candidate-based invariant generation in Sections~\ref{sec:loopinv} and~\ref{sec:cbinvariantgen}.  We then provide background on GPU kernels in Section~\ref{sec:gpukernels}, and give an overview of the GPUVerify tool for analyzing GPU kernels, explaining how GPUVerify incorporates candidate-based invariant generation, in Section~\ref{sec:gpuverifytool}.

\subsection{Inductive loop invariants}
\label{sec:loopinv}

A standard approach to reasoning about programs containing loops,
stemming from early work by Floyd~\cite{floyd1967assigning}, is to
apply a \emph{loop cutting transformation}, replacing each loop with a
loop-free sequence of statements that over-approximates the effects of
the loop on the program state.  We recap the process of reasoning about
a program via loop cutting, which is described in more technical detail elsewhere (e.g.~\cite{WeakestPrecondition05}).

To cut a loop, we need a property---a loop invariant---that holds each time the loop head is reached during execution.

\begin{figure}
\begin{minipage}[p]{0.30\columnwidth}
\begin{lstlisting}
while (c)
invariant $\phi$ {
  B;
}
\end{lstlisting}
\end{minipage}
\hfill
\begin{minipage}[p]{0.60\columnwidth}
\begin{lstlisting}
assert $\phi$;    // (base case)
havoc modset(B);
assume $\phi$;
if (c) {
  B;
  assert $\phi$;  // (step case)
  assume false;
}
\end{lstlisting}
\end{minipage}
\caption{%
The loop cutting transformation~\cite{WeakestPrecondition05} used by GPUVerify.  The loop on the left is transformed into the loop-free sequence of statements on the right.
}
\label{fig:loopcut}
\end{figure}

The transformation is depicted in Figure~\ref{fig:loopcut}: the input loop and its invariant $\phi$ on the left are turned into the loop-free sequence on the right.
In the transformed sequence, the loop invariant is checked in two places.
The first check ensures that the property is satisfied on entry to the loop (the base case), while the second check ensures the property is maintained by each execution of the loop body (the step case); if both can be proven, then the invariant is \emph{inductive}.
Observe that the step case requires the program to be in an arbitrary state which \emph{already} satisfies the loop invariant; this establishes the induction hypothesis.
The arbitrary state is obtained by first assigning a non-deterministic value to each variable possibly modified in the loop body (using the \code{havoc modset(B)} statement), and then assuming the loop invariant (using the \code{assume $\phi$} statement).
If the loop guard evaluates to \code{true}, the step case then checks that no assertions fail during execution of the loop body \code{B}, and that execution of the body results in a state that satisfies the loop invariant.

\subsubsection*{Example}

To illustrate inductive loop invariants, consider the annotated code snippet in Figure~\ref{fig:loopexample}, which repeatedly increments the variables $i$ and $j$.

\begin{figure}
\begin{lstlisting}
i := 0;
j := 0;

while (i $<$ 100)
  invariant $j = 2i$;
  invariant $j \le 200$;
{
  i := i + 1;
  j := j + 2;
}

assert $j = 200$;
\end{lstlisting}
\caption{An example code snippet annotated with loop invariants that allow proving the assertion $j = 200$.}
\label{fig:loopexample}
\end{figure}

The first invariant, $j = 2i$, is inductive in isolation.
The invariant holds trivially on entry to the loop when $i = j = 0$ (the base case), and is maintained by the loop body provided that the invariant and loop guard hold before the body executes (the step case):
\[
j = 2i\; \wedge\; i < 100 \Rightarrow (j+2) = 2(i+1) \,.
\]

The second invariant is \emph{not} inductive in isolation since $j \le 200$ and $i < 100$ do not imply $(j+2) \le 200$. However, the invariant is inductive in conjunction with $j = 2i$:
\[
j \le 200\; \wedge\; j = 2i\; \wedge\; i < 100 \Rightarrow (j + 2) \le 200 \, .
\]


The two invariants together with the negation of the loop guard suffice to prove the assertion near the bottom of Figure~\ref{fig:loopexample}:
\[
j = 2i\; \wedge\; j \le 200\; \wedge\; i \ge 100 \Rightarrow j = 200 \, .
\]

\subsection{Candidate-based invariant generation}
\label{sec:cbinvariantgen}

GPUVerify employs candidate-based invariant generation to compute inductive loop invariants automatically.
The technique speculates a finite set of potential invariants, called \emph{candidates}, that must be checked to ensure that they are, in fact, inductive invariants.
Checking is done by means of the Houdini algorithm~\cite{Houdini01}, which returns the unique, maximal conjunction of candidates that form an inductive invariant (see~\cite{DBLP:journals/ipl/FlanaganJL01} for a proof of this property).
The conjunction may be over the entire set of candidates (if all are proven to be inductive), but is more likely to be over a subset of these, due to some speculated candidates being false, or being true but not inductive.  In the worst case, the maximal conjunction returned by Houdini is over the empty set, meaning that the trivial invariant, \code{true}, is returned.

We provide some more details regarding the two phases of this approach.

\subsubsection{Phase one: the guess phase}
This phase supplies the candidates for a given program.
Guessing is domain specific and is free to use any static, dynamic, or hybrid technique.
A simple example is the use of syntactic checks that generate candidates based on pattern matching in the abstract syntax tree.
Importantly, this phase can be aggressive, generating a large set of candidates: false candidates cannot introduce unsoundness because the Houdini algorithm (in the check phase) will simply refute them.
Section~\ref{sec:rules} discusses the kinds of guesses performed by GPUVerify.

\begin{figure*}
\begin{minipage}[p]{0.25\textwidth}
\begin{lstlisting}
i := 0;
x := 1;
y := 2;
z := 3;

while (i < 10000)
  candidate $C_0: i = 0$;
  candidate $C_1: i \neq 0$;
  candidate $C_2: 0 \le i$;
  candidate $C_3: 0 < i$;
  candidate $C_4: i < 10000$;
  candidate $C_5: i \le 10000$;
  candidate $C_6: x \neq y$;
{
    temp := x;
    x := y;
    y := z;
    z := temp;
    i := i + 1;
}
\end{lstlisting}
\end{minipage}
\hfill
\begin{minipage}[p]{0.74\textwidth}
\begin{center}
\includegraphics[width=0.85\textwidth]{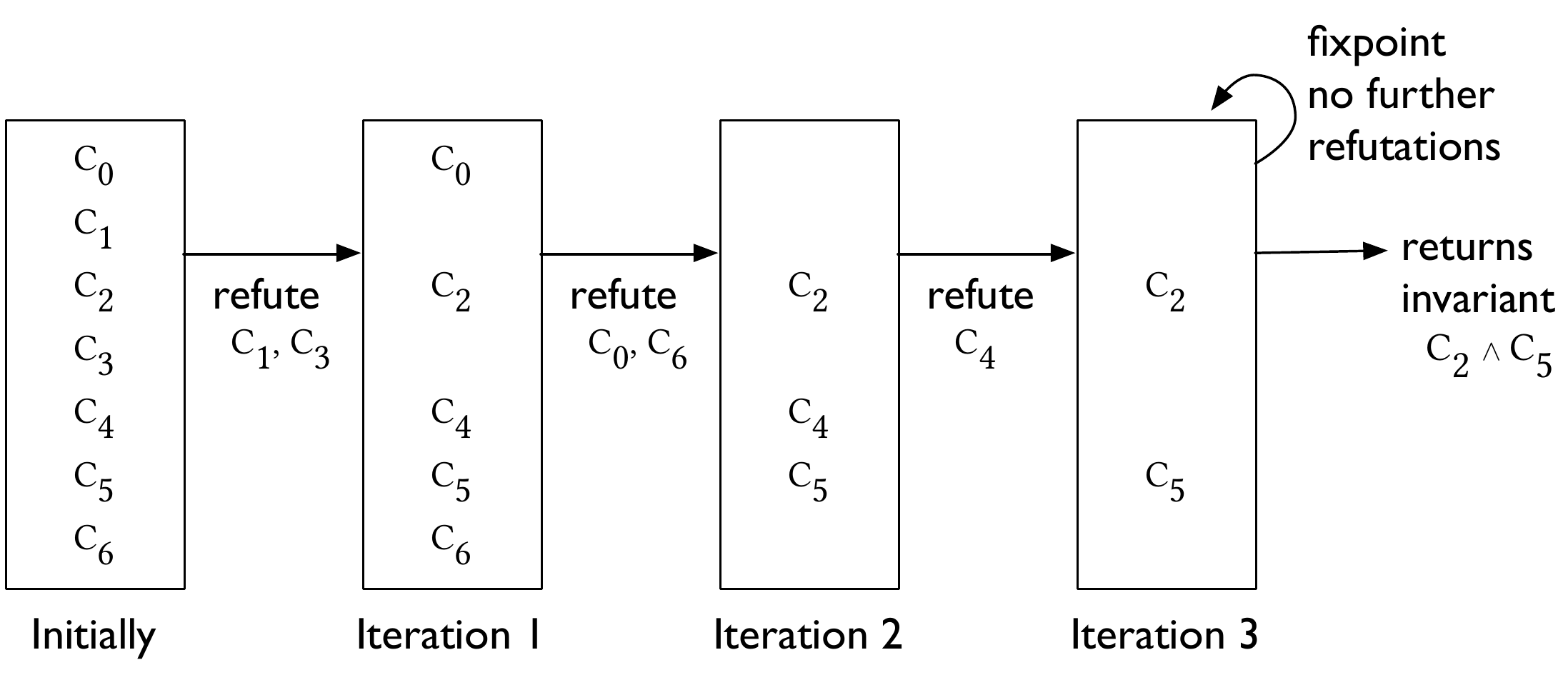}
\end{center}
\end{minipage}
\caption{An example program and a run of the Houdini algorithm, showing the candidates refuted at each iteration until a fixpoint is reached.}
\label{fig:houdiniexample}
\end{figure*}

\subsubsection{Phase two: the check phase}
Beginning with a full set of candidates, Houdini removes candidates that cannot be proven, until a fixpoint is reached.
A candidate may fail to be proven because the candidate is actually false, or because, despite being true, it cannot be proven to be inductive.  We refer to a candidate that Houdini removes as \emph{unprovable}, and say that Houdini \emph{refutes} such a candidate.  We say that a candidate that forms part of the maximal conjunction returned by Houdini is \emph{provable}.  Here, ``unprovable'' simply means that the invariant could not be proven to be inductive with respect to the current set of candidates.  If a candidate is true but unprovable, the candidate might become provable in the context of a larger set of candidates; this is because, as illustrated by the example of Section~\ref{sec:loopinv}, proving that an invariant is inductive can hinge on the availability of other supporting invariants.

Houdini is sound, deterministic, and terminating, assuming each call to the underlying SMT solver returns (which is not guaranteed if the logic used to encode verification conditions is undecidable).
For single-procedure programs, such as those analyzed by GPUVerify, the number of SMT solver calls is proportional to the number of candidates because Houdini only considers conjunctions of candidates.

Let us demonstrate Houdini using Figure~\ref{fig:houdiniexample}, which gives a program that repeatedly cycles the values $1,2,3$ around the variables $x,y,z$.
We assume the guess phase has speculated the candidates $C_0$ through $C_6$.
Houdini must now compute the maximal inductive invariant that is a conjunction of a subset of these candidates.
The figure shows how the set of candidates evolves during each iteration of the algorithm.
During the first iteration, Houdini removes $C_1$ and $C_3$ because they do not hold on loop entry (the base case).
No further candidates can be removed during this iteration: in the base case all other candidates hold, and the step case holds vacuously because candidates $C_0$ and $C_1$, which are mutually inconsistent, are both assumed to hold.
During the second iteration, the candidates $C_0$ and $C_6$ are refuted because they are not preserved by the loop.
To see why $C_6$ is not preserved, consider a state in which $x=1$ and $y=z=2$: this state satisfies $C_6$ on loop entry, but not after execution of the loop body.
During the third iteration, the candidate $C_4$ is refuted.
This candidate could not be removed until $C_0$ was removed since assuming $C_0$ allowed $C_4$ to be preserved by the loop.
This illustrates dependencies between candidates, where the refutation of a specific candidate is only possible after refutation of certain other candidates.
A fixpoint is reached during the final iteration: the remaining candidates, $C_2$ and $C_5$, form an inductive invariant, and Houdini returns $C_2\;\wedge\;C_5$.

It is worth noting that candidate $C_6$ is an invariant of the loop; it is refuted by Houdini because it is not inductive, as described above.  If the candidates $x \neq z$ and $y \neq z$ also had been provided initially then, because these candidates are mutually inductive with $C_6$, all three would have been returned by Houdini, in addition to $C_2$ and $C_5$.

\subsection{GPU kernels}
\label{sec:gpukernels}

A GPU \emph{kernel} is a program, typically written in CUDA~\cite{CUDA} or OpenCL~\cite{OpenCL}, that enables a general-purpose computation to be offloaded to a GPU.\footnote{Throughout the paper we present examples using notation from the CUDA programming model, though the kernels that we study are written in both CUDA and OpenCL.}
At run time, the kernel is launched with a thread configuration that specifies both the number of threads to run and the organization of the threads in blocks of size \code{blockDim}, where the blocks form a grid of size \code{gridDim} (both \code{blockDim} and \code{gridDim} may be multi-dimensional).
Each thread is parameterized by its thread and block identifiers (\code{threadIdx} and \code{blockIdx}), which allow it to compute memory addresses and make branch decisions unique to that thread.
Threads have access to thread-private memory and memory regions that are shared at the block and grid level.
Threads in the same block can communicate through shared memory and synchronize using \emph{barrier} operations.

GPU kernels are susceptible to \emph{data races} and \emph{barrier divergence}, which are programming errors.
A data race occurs when two different threads access the same location in shared memory, at least one access is a write, and there is no intervening synchronization.
Barrier divergence happens when threads reach distinct syntactic barriers or when the same barrier is reached under divergent control flow.
Various techniques have been proposed to assess whether GPU kernels are prone to, or free from, these errors.  We next describe the GPUVerify tool, which is the focus of study in this paper, and implements one of these techniques. We survey related approaches in Section~\ref{sec:related}.

\subsection{GPUVerify}
\label{sec:gpuverifytool}

\begin{figure}
\begin{center}
\includegraphics[width=\linewidth]{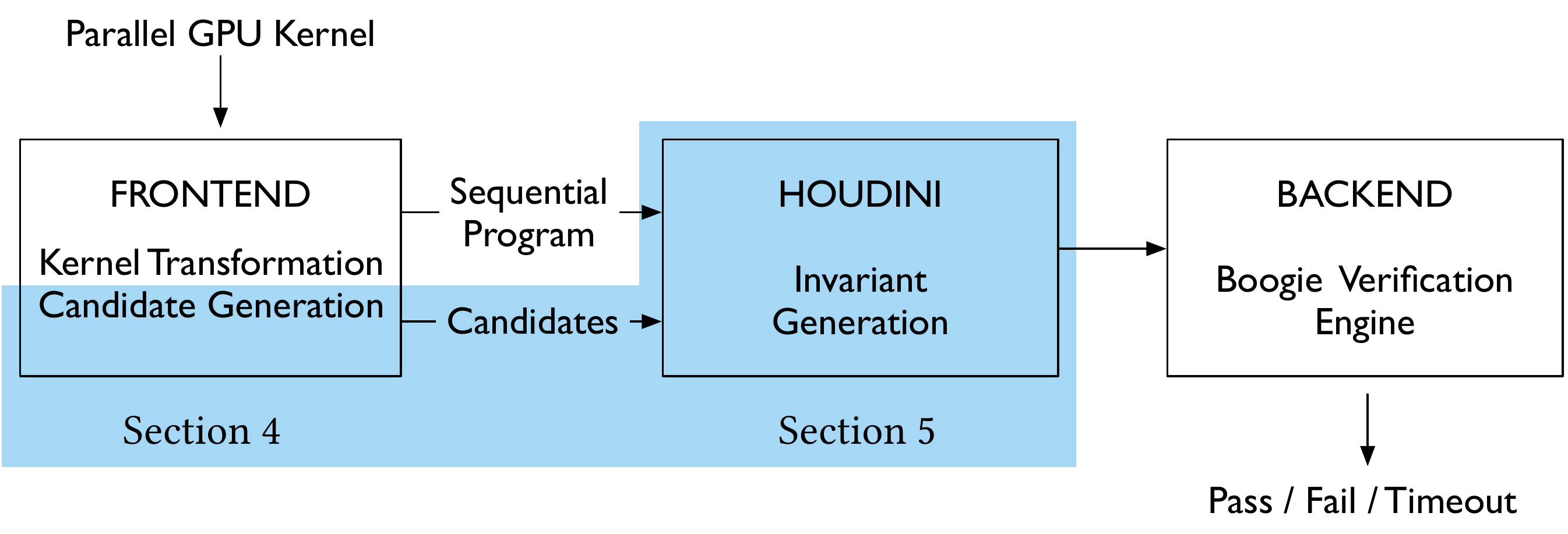}
\end{center}
\caption{The architecture of GPUVerify.  The shaded parts of the pipeline---candidate generation and accelerating Houdini---are the focus of this paper.}
\label{fig:gpuverify}
\end{figure}

The GPUVerify tool~\cite{GPUVerify12,GPUVerify13,TOPLAS15} has been
designed to automatically prove absence of data races and barrier
divergence for CUDA and OpenCL kernels.
The tool takes a kernel source file as input, and generates a
sequential program in the Boogie intermediate verification language~\cite{deline2005boogiepl}
annotated with automatically-generated assertions.  The translation
from kernel to Boogie program is performed in manner such that if the
Boogie program is free from assertion failures then the original kernel is
free from data races and barrier divergence.  The translation is
described elsewhere in a tutorial fashion~\cite{AVoCS14} and via a
formal operational semantics~\cite{GPUVerify13,TOPLAS15}.  It involves generating a
program that models the execution of a kernel by a pair arbitrary
threads according to a single, fixed schedule.  Instrumentation variables are used to detect races between the two threads, and
the possible effects of additional threads are over-approximated via abstraction.
The barrier-synchronous nature of the GPU programming model means that
proving correctness of the single-schedule sequential program for all
possible choices of thread identities suffices to show that the kernel
is free from data races and barrier divergence for all thread
schedules.  This allows well-established techniques for proving correctness of sequential programs to be leveraged in order to prove race- and divergence-freedom for highly parallel GPU kernels.

We note that GPUVerify performs full function inlining by default to increase verification precision.  In practice this is possible because recursion and function pointers are prohibited in OpenCL and rare in CUDA.

The architecture of GPUVerify is depicted in Figure~\ref{fig:gpuverify}.
The component labeled `FRONTEND' is responsible for transforming a GPU kernel into a corresponding sequential Boogie program,
and the `BACKEND' component uses the Boogie verification engine~\cite{Boogie} to verify the transformed kernel.
Boogie uses the loop cutting transformation discussed in Section~\ref{sec:loopinv} to abstract loops according to provided invariants, and calls upon an SMT solver, such as Z3~\cite{Z3} or CVC4~\cite{CVC4}, to check the resulting verification conditions that are generated.
The `HOUDINI' component of GPUVerify uses the Houdini algorithm, described in Section~\ref{sec:cbinvariantgen}, to automatically compute inductive loop invariants from a set of candidate invariants, which are speculated by GPUVerify's front-end.

GPUVerify is sound but incomplete: modulo bugs in the GPUVerify implementation, and a number of pragmatic assumptions made by the tool, a verified kernel is indeed guaranteed to be free from the sorts of defects that GPUVerify checks.  However, errors reported by the tool may be \emph{false positives}.
These spurious errors can arise due to:
\begin{itemize}
  \item abstract handling of floating-point operations,
  \item abstraction of the shared state, and
  \item insufficiently strong loop invariants.
\end{itemize}
In practice, we find that the last of these errors is the most common limitation of the tool, and the strength of the generated loop invariants is entirely governed by the effectiveness of our implementation of candidate-based invariant generation.

\section{Benchmark suite}
\label{sec:benchmarks}

We study invariant generation in GPUVerify using a set of $\SizeOfLOOPY$ benchmarks collected from nine sources:
\begin{itemize}
\item 54 OpenCL kernels from the \emph{AMD Accelerated Parallel Processing SDK} v2.6~\cite{AMDSDK},
\item 98 CUDA kernels from the \emph{NVIDIA GPU Computing SDK} v5.0 and v2.0~\cite{CUDASDK},
\item 16 CUDA kernels hand-translated from Microsoft's \emph{C++ AMP Sample Projects}~\cite{AMPSAMPLES},
\item 20 CUDA kernels originating from the \emph{gpgpu-sim} benchmarks~\cite{gpgpusim},
\item 15 OpenCL kernels from the \emph{Parboil} suite v2.5~\cite{Parboil},
\item 18 OpenCL kernels from the \emph{Rodinia} suite v2.4~\cite{Rodinia},
\item 50 OpenCL kernels from the \emph{SHOC} suite~\cite{SHOC},
\item 88 OpenCL kernels generated by the PPCG parallel code generator~\cite{PPCG} from the \emph{PolyBench/C} benchmarks v4.0a~\cite{PolybenchC}, and
\item 24 OpenCL kernels from the Rightware \emph{Basemark CL} suite v1.1~\cite{BasemarkCL}.
\end{itemize}
We refer to the above benchmarks as the \emph{\LOOPY set}.

All of the  suites are publicly available except for \emph{Basemark CL}, which was provided to us under an academic license.
The collection covers all the publicly available GPU benchmark suites of which we were aware at the start of our study, and we have made the versions of the kernels we used for our experiments available online.\footnote{\url{http://multicore.doc.ic.ac.uk/tools/GPUVerify/IEEE_TSE/}}

The kernel counts do not include \NumNotStudied kernels that we manually removed:
\begin{itemize}
  \item
    \NumVacuousPass kernels are trivially race- and divergence-free because they are executed by a single thread.
  \item
    \NumNotHandled kernels use either inline assembly, function pointers, thread fences, or CUDA \emph{surfaces}, which GPUVerify currently does not support.
  \item
    \NumRequireRefinements kernels are data-dependent (i.e.\ their control flow depends on array inputs to the kernel), which requires refinements of the GPUVerify verification method that cannot be applied automatically~\cite{OOPSLA13}.
  \item
    \NumNoLoops kernels are loop free and, hence, do not require invariant generation.
\end{itemize}

\subsection{Loop properties}

To discern the complexity of the kernels in the \LOOPY set, we counted the number of loops and the loop-nesting depth of each kernel after full inlining had been applied (recall from Section~\ref{sec:gpuverifytool} that GPUVerify performs full inlining by default).  Having many loops often makes a program hard to verify, and nested loops can be a particular challenge. In the case of a sequence of loops, proving that an invariant holds on entry to a given loop may only be possible if a sufficiently strong invariant is available for a preceding loop in the sequence; in the case where loop $L_2$ is nested inside loop $L_1$, invariants for $L_1$ may be required to prove that invariants of $L_2$ are inductive, and vice-versa.

We summarize the loop statistics in Table~\ref{tab:loopcount}.
The majority of kernels (68\%) only feature a single loop or a pair of (possibly nested) loops, but there are still a significant number of kernels (122) with a larger number of loops.
At the very extreme, the \code{heartwall} kernel from the Rodinia suite features 48 loops that are syntactically distinct, i.e.\ they do not arise as a result of inlining.

Nested loops occur in 37\% of kernels.
Deep loop-nests are mostly found in the PolyBench/C kernels, with 42 of those kernels having a maximum loop-nest depth of three, and 9 having a maximum loop-nest depth of four. All kernels with a maximum loop-nest depth of five also originate from this set.

\begin{table}
\centering
\caption{Basic loop statistics of the \LOOPY set.}
\begin{tabular}{l|rrrrrrr}
\textbf{Number of loops} & 1 & 2 & 3 & 4 & 5 & 6 & 7+ \\
\toprule
Kernels & 168 & 93 & 52 & 34 & 19 & 2 & 15\phantom{+} \\
\end{tabular}
\vskip\baselineskip
\begin{tabular}{l|rrrrr}
\textbf{Maximum loop-nest depth} & 1 & 2 & 3 & 4 & 5 \\
\toprule
Kernels & 243 & 67 & 55 & 14 & 4 \\
\end{tabular}
\label{tab:loopcount}
\end{table}

\subsection{Obtaining scalar kernel parameters}
\label{sec:methodology:params}

Most kernels in the \LOOPY set are designed to be race free only for \emph{constrained} thread configurations and input values.
These preconditions are often implicit and very rarely documented, and any information that does exist appears as informal source code comments.
Unfortunately, suitable preconditions must be communicated to GPUVerify in order to avoid spurious error reports.

We were able to confidently add general preconditions to some kernels by hand---kernels that we were familiar with, or that were well-documented or sufficiently simple to understand at a glance.  However, for the majority of kernels we solved the above problem by discovering constraints for kernel \emph{scalar} parameters in the following way:
\begin{enumerate}
\item We ran the application in which the kernel was embedded and intercepted kernel calls using dynamic library instrumentation to note the input parameters of each kernel.
For OpenCL applications we used KernelInterceptor~\cite{KernelInterceptor}.
For CUDA applications we used a similar prototype tool.\footnote{\url{https://github.com/nchong/cudahook}}
Running the kernels in interception mode led to an intercepted value for every kernel input parameter.
\item We ran GPUVerify on the kernel in bug-finding mode, assuming the observed thread configuration but with unconstrained formal parameters.
In bug finding mode, GPUVerify unrolls loops up to a fixed depth of two to avoid the need for loop invariants, thereby reducing the possibility of false positives.
If GPUVerify reported a possible data race, and if on manual inspection of the race we concluded that an unconstrained \emph{integer} parameter contributed to the conditions causing the race, we added a precondition to constrain the value of the parameter to the value intercepted for it.
This step was repeated until GPUVerify was satisfied; in the extreme case this led to constraining all integer parameters.
\end{enumerate}

We only added integer preconditions, because GPUVerify handles floating-point and array data abstractly. It is atypical for race freedom to require preconditions on floating-point inputs, and cases where race freedom is dependent on array preconditions requires manual methods (see the earlier discussion regarding kernels removed from our study).

The above process offers a pragmatic solution to garner a suitable but not overly constrained precondition, although the most general precondition may be missed.
For example, in the case of the matrix transpose kernel of Figure~\ref{fig:transpose} (to be discussed in Section~\ref{sec:accessbreaking}), the process led to the precondition $\code{height} = 8$ being added for an $8 \times 8$ input matrix, instead of the more general $\code{height} = \code{gridDim.y} \times \code{TILE\_DIM}$, which would allow us to also prove the kernel correct for matrices of different sizes.
Another common case is for a GPU kernel to require a scalar argument to have a value that is a power of two within a particular range.  Our interception-based approach would constrain such a parameter to a particular power of two used at runtime, restricting subsequent analysis of the kernel to that particular value.

We acknowledge that the form of kernel preconditions can have an impact on
invariant generation: a strong precondition might facilitate the use
of strong, easy-to-infer loop invariants, while a weaker precondition
might necessitate more general, harder-to-infer invariants.  Nevertheless, equipping our kernels with a pragmatically-obtained set of preconditions still led to a challenging set of benchmarks for invariant generation.

After completing our study we reviewed the number of preconditions inserted using the above methodology (see Table~\ref{tab:preconditions}).
We found that 61\% of kernels required preconditions, and that on average one precondition was required.
The largest number of preconditions required for a single kernel was 26 (for the \code{heartwall} kernel from the Rodinia suite).

\begin{table}
\centering
\caption{Number of introduced preconditions.}
\begin{tabular}{l|rrrrrrrrr}
\textbf{Preconditions} &   0 &  1 &  2 &  3 &  4 & 5 & 6 & 7 & 8+ \\
\toprule
Kernels                & 151 & 90 & 86 & 37 & 11 & 5 & 0 & 2 & 1\phantom{+} \\
\end{tabular}
\label{tab:preconditions}
\end{table}

\section{Candidate generation in GPUVerify}
\label{sec:rules}

We now explain how we devised the rules that generate candidates in
GPUVerify, which was driven by the aim of automatically verifying as many of our benchmark programs as possible.  We also propose several new metrics for evaluating candidate-based invariant generation, and evaluate our rules across our benchmark set using these metrics.
In Section~\ref{sec:candidategenerationprocess} we describe
our strategy for devising new
candidate generation rules.  In Section~\ref{sec:accessbreaking} we discuss, as an
example, the intuition behind one of the rules that we developed (the remaining rules are outlined in Appendix~\ref{app:rules}).  In
Section~\ref{sec:evaluationofrules}, we assess the
effectiveness of, and relationship between, the rules that we devised.  During this process we discovered a number of defects in the GPU kernels we studied.  We briefly document these issues in Section~\ref{sec:bugs}.

\subsection{Process for deriving candidate-generation rules}\label{sec:candidategenerationprocess}

To gauge the need for invariant generation in GPUVerify to prove the absence of data races and barrier divergence, we attempted to verify each kernel in the \LOOPY set without loop invariants being either manually or automatically supplied.
If a kernel verifies under these conditions then we say the kernel is \emph{trivial}; typically, a trivial kernel has loops that either do not access the shared state at all, or that access shared arrays that are never written to by the kernel.
We found $\SizeOfNonTrivial$ out of $\SizeOfLOOPY$ kernels (65\%) to be \emph{non}-trivial, the majority of the \LOOPY set.
Hence, assuming these kernels are indeed correct (and can be verified as such via suitable invariant annotations), invariant generation is crucial to the precision of GPUVerify.

The set of non-trivial kernels facilitated the design of new rules as follows:
\begin{enumerate}
  \item We randomly picked a small number of kernels from the set, and manually determined a minimal set of loop invariants that enabled their verification.

  \item Each picked kernel was updated to include the necessary loop invariants as \emph{user-defined invariants}.

  \item Common patterns were identified among the user-defined invariants that might apply in a wider setting. For each such pattern, we introduced a candidate-generation rule to GPUVerify.

  \item\label{item:removesubsumed} We removed any user-defined invariants that were subsumed by the introduced rules.
\end{enumerate}



We iterated the above process of invariant discovery until all kernels in the \LOOPY set---bar five kernels which we found to contain data races---could be verified automatically through a combination of invariants generated by our candidate-based approach and invariants provided manually (or, in the case of PolyBench/C, and as explained below, generated by a compiler).
We rigorously applied step~\ref{item:removesubsumed} above to ensure that all remaining user-defined invariants were necessary in order for verification to succeed using the candidate invariants generated by GPUVerify, ensuring the removal of any manually supplied invariants that were originally necessary but subsequently subsumed by generated invariants.

The sketched process led to the development of \NumRules rules, summarized in Appendix~\ref{app:rules}.  In
Section~\ref{sec:accessbreaking} we explain how one such rule was developed.

\subsubsection*{Compiler-generated invariants for the PolyBench/C suite}
The 88 kernels from the PolyBench/C suite were generated by PPCG~\cite{PPCG}, a polyhedral compiler equipped with a back-end that can compile polyhedral code, and some extensions thereof~\cite{PENCIL}, into OpenCL.

Many of the machine-generated kernels feature a large number of
loops, in some cases deeply nested due to the application of loop
tiling program transformations.  We verified a selection of these
kernels by manually working out sufficient invariants, and found these
invariants to be divided into two sorts: basic invariants about loop
bounds (similar to the invariants required by many other
kernels) for which we had already devised suitable
candidate-generation rules, and intricate, specialized invariants
related to the memory access patterns associated with polyhedral code
generation.  For the latter invariants we worked with the lead
developer of PPCG to add an option whereby the compiler can automatically
generate a program-wide invariant characterizing the access patterns of
all loops.  Adding the option was possible by virtue of the rich information
regarding memory access patterns available in PPCG, which it uses to
perform code generation.  The invariants generated by the compiler are independently checked by GPUVerify.  For more details see~\cite{CARPDeliverable}.

By combining the compiler-generated invariants regarding access patterns with invariants inferred by GPUVerify relating to loop bounds it was possible to verify almost all PolyBench/C kernels.  Because the compiler-generated invariants are specific to this class of kernels, and because generation of the invariants by the compiler is reliable, we decided not to propose candidate-generation rules to speculate these invariants automatically. There were four kernels that did not verify out-of-the-box using this strategy. Each of these kernels required invariants unique to that kernel and, hence, we opted to supply these invariants manually.

\subsection{Access breaking rule}\label{sec:accessbreaking}

As an illustrative example, we now describe a memory access pattern
that we observed in a variety of the kernels, outline the process
by which we manually derived invariants to characterize the access pattern,
and comment on how these led to a candidate-generation rule that automates
our manual process.

\begin{figure*}
\begin{minipage}[b]{0.58\textwidth}
\begin{lstlisting}[linewidth=\linewidth]
#define TILE_DIM 4
#define BLOCK_ROWS 2

__global__ void transpose(float *odata, float *idata,
                          int width, int height) {
  int xIndex = blockIdx.x * TILE_DIM + threadIdx.x;
  int yIndex = blockIdx.y * TILE_DIM + threadIdx.y;

  int index_in  = xIndex + width  * yIndex;
  int index_out = yIndex + height * xIndex;

  for (int i = 0; i < TILE_DIM; i += BLOCK_ROWS) {
    odata[index_out + i] = idata[index_in + i * width];
  }
}
\end{lstlisting}
\end{minipage}
\hfill
\begin{minipage}[b]{0.39\textwidth}
\includegraphics[width=\linewidth]{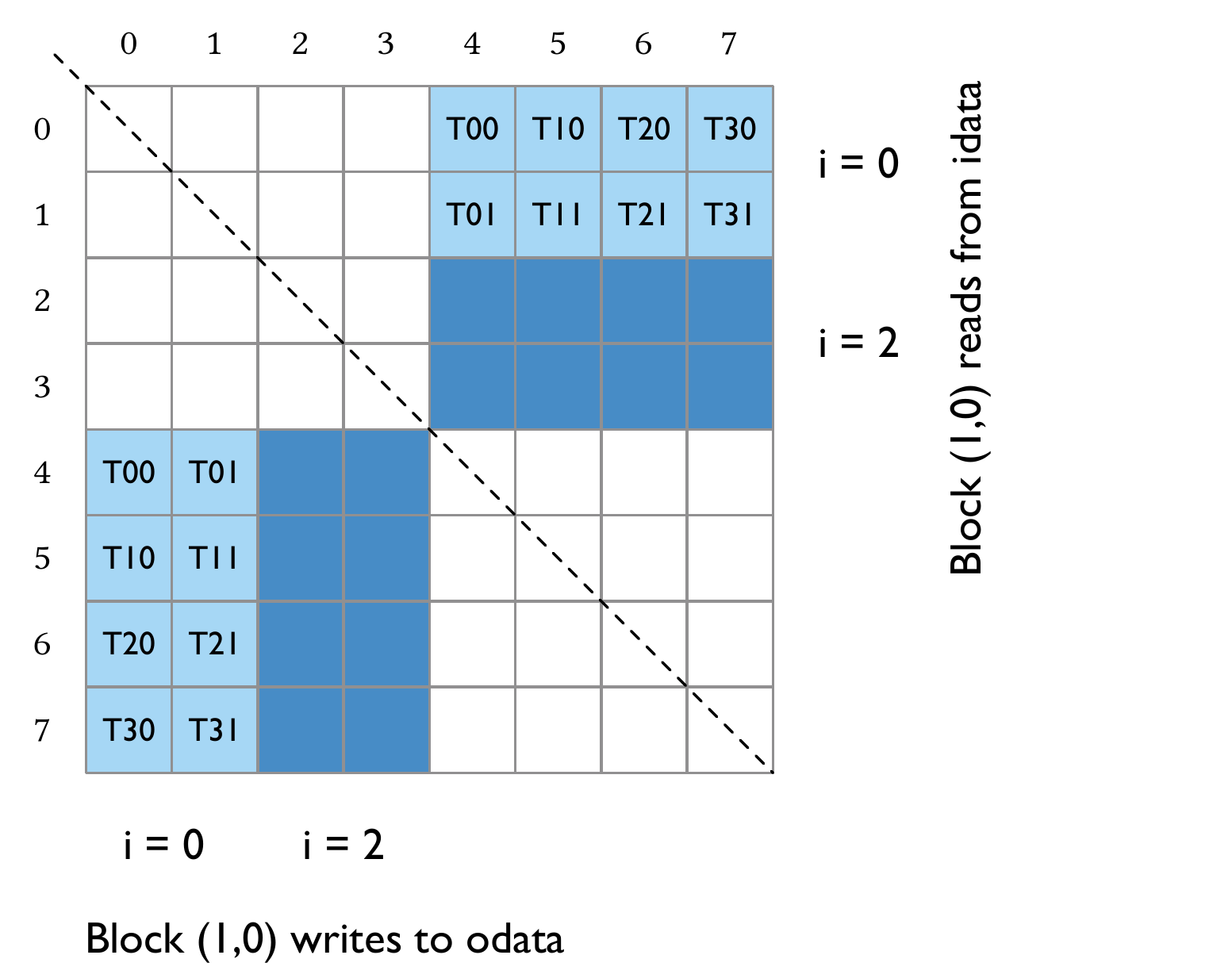}
\end{minipage}
\caption{A matrix transpose example taken from the CUDA SDK.  The right side depicts reads of \code{idata} and writes of \code{odata} by block $(1,0)$ for $\code{i} = 0$ and $\code{i} = 2$. Note that the \code{idata} and \code{odata} arrays are disjoint.}
\label{fig:transpose}
\end{figure*}

Consider the matrix transpose kernel of Figure~\ref{fig:transpose}, which is taken from the CUDA SDK~\cite{CUDASDK}.
The kernel reads from an input matrix \code{idata} of dimension \code{width} by \code{height} and writes to an output matrix \code{odata}.
The matrices are stored in row-major order, meaning that an element $A_{x,y}$ of a matrix $A$ is stored in a linear array at offset $x + \code{width} \times y$.
The kernel is invoked with a 2-dimensional grid of 2-dimensional thread blocks, with each block of size $\code{TILE\_DIM} \times \code{BLOCK\_ROWS}$.
Each block is assigned a square tile of dimension \code{TILE\_DIM} of the input and output matrices.
Individual threads within a block stride along the assigned tile in increments of \code{BLOCK\_ROWS}.
During each iteration of the loop, each block of threads copies $\code{TILE\_DIM} \times \code{BLOCK\_ROWS}$ elements from \code{idata} to \code{odata}.
For example, if the matrix dimension is $8 \times 8$, with $\code{TILE\_DIM} = 4$ and $\code{BLOCK\_ROWS} = 2$, then the kernel is invoked with a $2 \times 2$ grid of $4 \times 2$ blocks, and each block is assigned a tile of $4 \times 4$ elements.
The read and write assignment of block $(1,0)$ is shown on the right side of Figure~\ref{fig:transpose}.
In the figure, we see e.g.\ that thread $(1,1)$ of block $(1,0)$ assigns $\code{idata}_{5,1}$ to $\code{odata}_{1,5}$ and $\code{idata}_{5,3}$ to $\code{odata}_{3,5}$ when $i$ equals $0$ and $2$, respectively.

Intuitively, GPUVerify checks for data race-freedom by analyzing the intersection of read and write sets of all distinct thread pairs.
In this example, the kernel is free from data races since \code{idata} is only ever read from, and distinct threads write to distinct offsets of \code{odata}.
The loop invariants that we require must summarize the writes to \code{odata}.
If $W_t$ denotes the set of all writes that a thread $t$ has issued, then a set of invariants that relates $W_t$ to its thread identifiers is useful because two distinct threads must always have at least one block or thread identifier that is distinct:

{
\footnotesize
\[
\begin{array}{r@{\ }c@{\ }c@{\ }c@{\ }c@{\ }l@{\quad}c@{\quad}l}
\forall w \in W_{t}. &
  ((w &  / & \code{height}) &  / & \code{TILE\_DIM}) & = & \code{blockIdx.x} \\
\wedge \ &
  ((w &  / & \code{height}) & \% & \code{TILE\_DIM}) & = & \code{threadIdx.x} \\
\wedge \ &
  ((w & \% & \code{height}) &  / & \code{TILE\_DIM}) & = & \code{blockIdx.y} \\
\wedge \ &
  ((w & \% & \code{height}) & \% & \code{TILE\_DIM}) \mathrlap{\ \%\ \code{BLOCK\_ROWS}} \\
&
      &    &                &    &                   & = & \code{threadIdx.y}
\end{array}
\]
}

The invariant is trivially satisfied on loop entry, because on entry $W_t$ is empty for all threads.
To see that the invariant is maintained by the loop consider an arbitrary write $w$ from thread $t$ to \code{odata}.
The access will be of the form $\code{index\_out} + \code{i}$, where \code{i} is a multiple of \code{BLOCK\_ROWS}.
In other words, $w$ is of the form:

{
\footnotesize
\[
\begin{array}{l}
(\code{blockIdx.x} \times \code{height} \times \code{TILE\_DIM}) \\
\qquad + \ \ (\code{threadIdx.x} \times \code{height}) \\
\qquad + \ \ (\code{blockIdx.y} \times \code{TILE\_DIM}) \\
\qquad + \ \ \code{threadIdx.y} \\
\qquad + \ \ \code{i}
\end{array}
\]
}

\noindent
which shows that the invariant is indeed maintained.

We refer to the above type of invariant as \emph{access breaking} since the access pattern is broken down into the components identifying thread $t$.
We have derived a candidate-generation rule to speculate access breaking invariants.  For each memory access appearing in a kernel, we use cheap abstract syntax tree-based pattern-matching to determine whether the access uses thread identifiers.
If so, we trigger access breaking, which consists of rewriting the access expression to generate a possible equality for each of the components identifying a thread.
%
%

The precise conditions under which GPUVerify triggers access breaking, and the particular candidates that this rule generates, are intricate and were devised in an example-driven manner; for exact details please refer to the \texttt{RaceInstrumenter} class in the GPUVerify source code, and search for the \texttt{accessBreak} tag.

\subsection{Evaluation of rules}\label{sec:evaluationofrules}

GPUVerify presently has \NumRules rules for generating can\-di\-dates---see Appendix~\ref{app:rules} for a short description of each of these rules.
Here we evaluate the effectiveness of the rules.
In particular, we aim to provide answers to the following questions:
\begin{itemize}
  \item \textbf{Rule generality:} Do the rules cause candidate invariants to be generated for a variety of kernels?
  \item \textbf{Rule hit rate:} To what extent do the rules produce provable candidates?
  \item \textbf{Rule worth:} How often are provable candidates generated by a rule \emph{essential} for precise reasoning?
  \item \textbf{Rule influence:} To what extent are the rules independent, in terms of provability of the candidates they generate?
  \item \textbf{Increase in precision:} For how many kernels does candidate-based invariant generation make the difference between verification succeeding or failing?
\end{itemize}

This set of questions, and our systematic approach to answering them experimentally, is one of the main contributions of our work: the questions and methods easily generalize beyond our domain of GPU kernel verification, thus we believe they provide a framework that will be useful to other researchers interested in designing and evaluating candidate-based approaches to invariant generation.

\subsubsection{Experimental setup}
\label{sec:expsetup}

The experiments in this section were conducted on a machine with a 2.4GHz 4-core Intel Xeon E5-2609 processor and 16GB of RAM, running Ubuntu 14.04 and using CVC4 v1.5-prerelease, Clang/LLVM v3.6.2, and Mono v3.4.0.
In GPUVerify, we enabled user-defined invariants and set the timeout to 30~minutes.
We chose a large timeout to allow more kernels to be explored, thus yielding a larger evaluation.
We enabled user-defined invariants to ensure that, with all candidate generation rules enabled, each kernel would verify.  This was necessary to meaningfully measure rule \emph{worth} and rule \emph{influence} in Sections~\ref{sec:ruleworth} and~\ref{sec:ruleinfluence}, and reflects the current state of GPUVerify where complex examples require a combination of manually-supplied and automatically-inferred invariants.

We removed 12 kernels from experiments reported upon in this section---unless indicated otherwise---leav\-ing 372 kernels from the \LOOPY set. In the case of 11 of the 12 kernels, GPUVerify did not complete, even with the generous timeout of 30 minutes.
In the case of one kernel GPUVerify ran out of memory.
Among these kernels, the timeouts and memory-out appear to be due to: large invariants generated by the PPCG (5 \emph{Polybench/C} kernels);
loops that contain many accesses, leading to a large number of access-related candidates being generated and significant reasoning required to prove race-freedom (4 \emph{gpgpu-sim} kernels, plus one \emph{SHOC} kernel, which is the kernel that exhausts our memory limit); the presence of a very large number of loops (the \emph{Rodinia} \texttt{heartwall} kernel); loops with an iteration space that leads to hard-to-prove invariants being generated by the loopBound rule (1 kernel from the \emph{NVIDIA GPU Computing SDK} v5.0, which in fact verifies when no invariants are speculated).

With enough time and computing resources, we could have included these kernels in our evaluation of the new candidate generation rules.  We decided not to so that our experiments would complete within a feasible time budget.  Relatedly, we could have included the 82 race- and divergence-free kernels that we eliminated from our test set as described in Section~\ref{sec:benchmarks}, since our candidate generation rules would still potentially trigger on these benchmarks.  However, we preferred to restrict our evaluation to multi-threaded examples where there really is potential for concurrency-related problems.

\subsubsection{Experiment: rule generality}

Our first hypothesis was that the conditions under which a rule triggers would be found in a variety of non-trivial kernels but in few, if any, trivial kernels.
To test this, we recorded, for each rule, the number of trivial and non-trivial kernels that contained at least one candidate produced by that rule.

\begin{table}
\centering
\caption{The number of kernels for which each rule triggers, and the hit rate and essentiality of each rule.}
\label{tab:rules:generality}
\label{tab:rules:hitrate}
\label{tab:precision:rules}
\begin{tabular}{l|rrr|r|r}
  & \multicolumn{3}{|c|}{Kernels triggering a rule}
  & \\
Rule & Non-trivial & Trivial & Total & Hit rate & Essentiality \\
\toprule
r0 & 70 & 0 & 70 & 74\% & 36 \\
r1 & 204 & 7 & 211 & 89\% & 88 \\
r2 & 30 & 0 & 30 & 83\% & 3 \\
r3 & 30 & 0 & 30 & 60\% & 3 \\
r4 & 85 & 7 & 92 & 59\% & 0 \\
r5 & 141 & 6 & 147 & 99\% & 1 \\
r6 & 81 & 3 & 84 & 100\% & 3 \\
r7 & 46 & 0 & 46 & 91\% & 32 \\
r8 & 184 & 6 & 190 & 75\% & 0 \\
r9 & 238 & 17 & 255 & 47\% & 30 \\
r10 & 143 & 6 & 149 & 87\% & 113 \\
r11 & 46 & 0 & 46 & 91\% & 4 \\
r12 & 103 & 0 & 103 & 49\% & 45 \\
r13 & 114 & 2 & 116 & 52\% & 36 \\
r14 & 50 & 5 & 55 & 40\% & 16 \\
r15 & 50 & 5 & 55 & 29\% & 0 \\
r16 & 46 & 0 & 46 & 4\% & 1 \\
r17 & 14 & 0 & 14 & 4\% & 10 \\
r18 & 7 & 0 & 7 & 90\% & 1 \\
\end{tabular}
\end{table}

Columns 2--4 of Table~\ref{tab:rules:generality} displays the results.
We see that the \SizeOfTrivial trivial kernels rarely contain patterns that trigger a rule. This is positive because trivial kernels can be verified without the provision of any invariants, and checking the provability of superfluous candidates is likely to slow down the verification process.
Overall, most rules are neither too speculative (rule r9 is activated by 255 kernels, the maximum) nor too bespoke (rule r17 is activated by 14 kernels, the minimum).\footnote{We disregard rule r18 here, because the rule is tied to a particular GPUVerify command-line option that turns on an extension for awareness of warp-level synchronization in CUDA~\cite{EthelNFM}. The extension is only enabled for the 9 kernels in the \LOOPY set.}
We believe this confirms that the process by which we introduced rules into GPUVerify has merit.

\subsubsection{Experiment: rule hit rate}

We further conjectured that a reasonable number of candidates produced by a rule would be provable.
We scrutinized this hypothesis by counting the following for each kernel in our evaluation set:
\begin{itemize}
  \item the number of candidates produced by a rule, and
  \item the split for these candidates between those that are provable and unprovable.
\end{itemize}
The \emph{hit rate} of a rule is then the percentage of candidates that are provable.

The results appear in the fifth column of Table~\ref{tab:rules:hitrate}.
One rule (r6) has a hit rate of 100\%, and five rules (r1, r5, r7, r11, r18) have hit rates close to 100\%.
We did not expect to find this many rules with such high hit rates because we designed our rules to guess candidates aggressively, in general preferring to err on the side of generating a large number of candidates (of which many may turn out to be unprovable) to increase the chance of generating a provable candidate in a scenario where it is needed for verification to succeed.
Conversely, we see that rules r16 and r17 speculate poorly, indicating that the conditions under which they trigger should be refined.
However, rule r17 is in danger of becoming too specialized, as it is already unsuccessful at producing candidates for many kernels; the rule only triggers in 14 cases.

\subsubsection{Experiment: rule worth}
\label{sec:ruleworth}

We cannot conclude from the previous experiment that rules with high hit rates are beneficial to verification.
A devious rule can generate trivially provable yet useless candidates for any kernel.
Hence, we wanted to know whether rules produce constructive candidates that actually enable verification.

Our hypothesis was that there would be numerous kernels whose verification depended on the candidates generated by a particular rule, given that we engineered rules in response to non-verifying kernels.
We tested this by observing whether GPUVerify reports differently for a kernel after a rule had been disabled.
Specifically, we say that rule $r$ is \emph{essential} for a kernel if two conditions are satisfied:
\begin{enumerate}
 \item the kernel verifies when \emph{all} rules are enabled, and
 \item disabling all candidates generated by rule $r$ causes verification to fail or time out.
\end{enumerate}

We counted the number of kernels for which each rule is essential.
The results are shown in the final column of Table~\ref{tab:precision:rules}.
The sum of the ``essentiality'' column is 422, meaning that there are 422 distinct (kernel, rule) pairs where a rule is essential for a kernel. Note that multiple rules may be essential for the same kernel.

At a glance, it may seem odd that rule r17 triggers for 14 kernels---and is essential for 10 of these---and yet only has a hit rate of 4\%.
The low hit rate is due to it being a measure of the generated candidates that are provable invariants, and the particular rule generating many candidates per kernel.

Unessential rules (r4, r8, r15) are redundant and could be removed without affecting the precision of GPUVerify on our benchmarks: they have been superseded by more general rules.  It is possible that removal of these rules could change the \emph{performance} of GPUVerify.  This is because, for a given program, there are typically many ways to phrase a suitable set of invariants for verification, and the way the invariants are phrased can affect the ease or difficulty with which an underlying solver can process the resulting verification conditions.

\subsubsection{Experiment: rule influence}\label{sec:ruleinfluence}

Our final conjecture was that the rules were independent of each other, having been designed largely in isolation.
To investigate this hypothesis, we observed whether disabling a rule in GPUVerify affected the hit rate of any of the other rules.
In this case, we say that the disabled rule \emph{influences} the rule whose hit rate changed.

Observe that if rule $r$ influences rule $s$, then the hit rate of $s$ can only decrease when r$x$ is disabled, because Houdini returns the unique, maximal conjunction of candidates forming an inductive invariant.
To see this, consider the example from Section~\ref{sec:loopinv}.  The second invariant, $j \le 200$, is only inductive in conjunction with the first, $j = 2i$.
Hence, not speculating the first makes proving the second impossible.

\begin{figure}
\includegraphics[width=\linewidth]{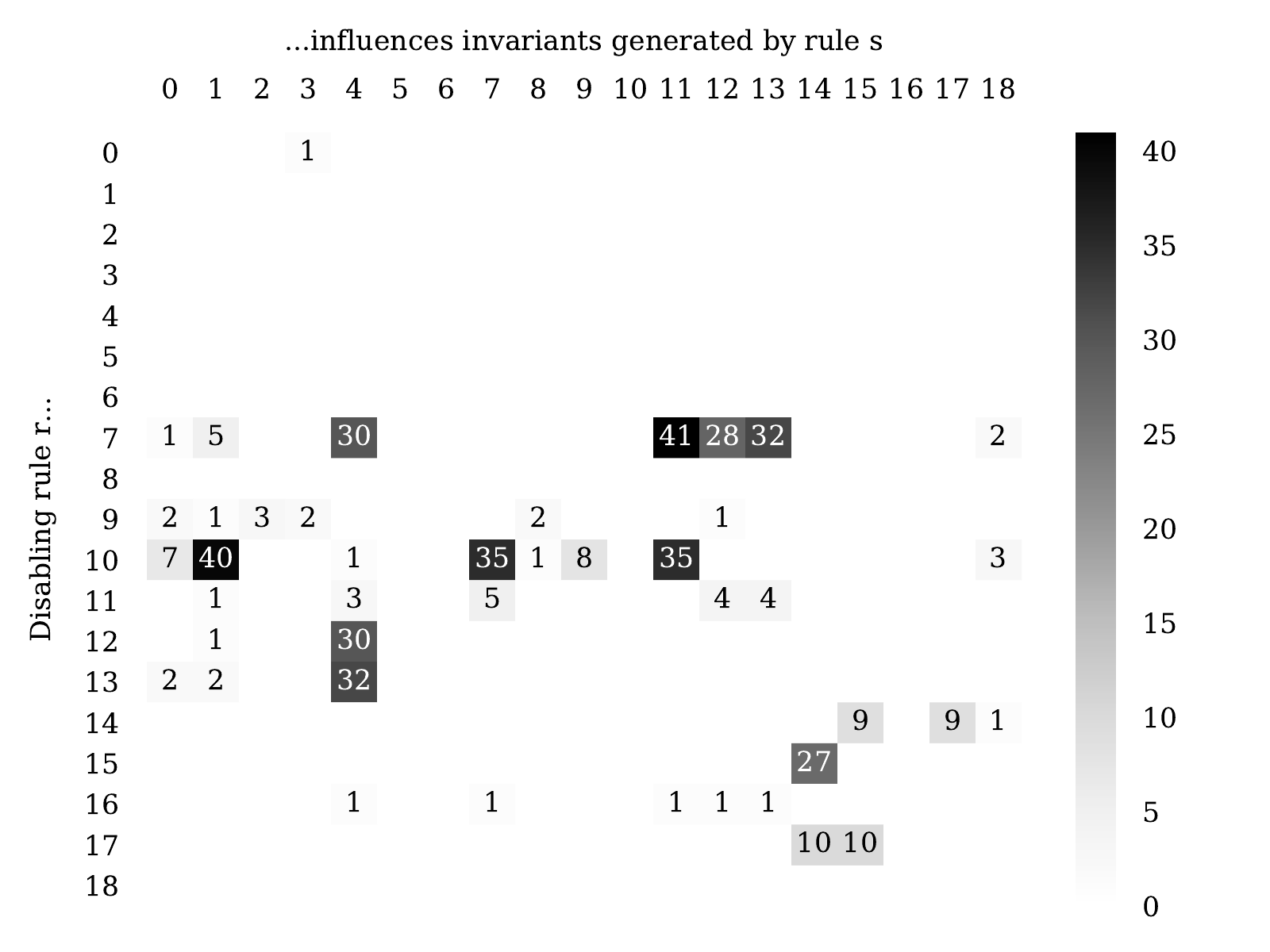}
\caption{For each pair of rules $(r,s)$ the heatmap shows the number of kernels where rule $r$ influenced rule $s$.  The color bar at the right of the figure indicates the number of kernels associated with the colors used in the heatmap.}
\label{fig:influencing}
\end{figure}

The results of disabling each rule in turn in GPUVerify are represented by the heat map of Figure~\ref{fig:influencing}.
For each pair of rules $(r,s)$ we give the number of kernels where $r$ influenced $s$.
We see that the matrix is relatively sparse, with only 43 non-zero entries, which demonstrates that most rules do not influence each other when applied to our benchmarks.
The major exceptions are rules r7 and r10, which speculate fundamental facts
related to loop counters; these loop counters are likely to be used to index into shared arrays.

\subsubsection{Experiment: overall increase in precision}

An unanswered question is how much more precision is afforded by the rules as a whole.
To answer this, we launched GPUVerify with all rules disabled and then with all rules enabled, and we counted the number of kernels that verified.
Disabling all rules, $\SizeOfTrivial$ trivial kernels and 0 non-trivial kernels verified, whereas with all rules enabled, $\SizeOfTrivialVerifiedWithCandidates$ trivial kernels and $\SizeOfNonTrivialVerified$ non-trivial kernels verified, a net gain of $\NetGainWithCandidates$ kernels.
The slight drop in the number of trivial kernels verified is caused by a timeout: the rules hinder performance.  In Section~\ref{sec:performance} we turn our attention to this problem.

\subsection{Defects detected during the process of deriving candidate-generation rules}\label{sec:bugs}

The process of attempting to prove data race- and barrier divergence-freedom of a large set of kernels led us to discover that a number of the kernels we considered were in fact incorrect.  We also found two bugs in the PPCG compiler discussed in Section~\ref{sec:candidategenerationprocess}.  We briefly detail the defects and our efforts to report the issues in order for them to be fixed:

\begin{itemize}
\item
A missing barrier in the SHOC \emph{sort} \verb!top_scan! kernel causing a data race. This race was reported and subsequently fixed.\footnote{\url{https://github.com/vetter/shoc/issues/30}}
\item
 Two threads writing to the same array location in the CUDA 5.0 \emph{convolutionFFT2D} \verb!spPostprocess2D! kernel. This race was reported to Nvidia and is fixed in CUDA 7.0.
\item
A missing barrier in the Rodinia \emph{SRAD} \verb!reduce! kernel between the initialization and use of an array. We reported this issue, and it has been fixed in version 3.1 of the suite.\footnote{See acknowledgment at \url{https://www.cs.virginia.edu/~skadron/wiki/rodinia/index.php/TechnicalDoc}}
\item
Overlapping writes to an array due to an incorrectly set kernel parameter in the Rodinia \emph{kmeans} \verb!kmeans_swap! kernel. This issue has also been fixed in version 3.1 of the suite, in response to our report.
\item
Two threads writing to the same array location in the Rodinia \emph{leukocyte} \verb!dilate! kernel.  This has also been fixed in version 3.1 of the suite, in response to our report.
\item
A similar issue in the Rodinia \emph{particle filter} \verb!normalize_weights_single! kernel, which has been reported and confirmed, but is not yet fixed.
\item
A data race due to an incorrectly initialized vector in the Parboil \emph{cutcp} benchmark, more specifically in the \verb!opencl_cutoff_potential_lattice! kernel, which we have reported, with confirmation awaiting.
\item
A data race affecting several tests in the SHOC \emph{DeviceMemory} benchmark, which we have reported.\footnote{\url{https://github.com/vetter/shoc/issues/31}} However, because these tests are performance micro-benchmarks that use random data, the race may not be regarded as important.
\item
Data races in some of the kernels generated by PPCG due an issue related to PPCG's use of schedule trees~\cite{ScheduleTree} and due to PPCG accidentally ignoring shared arrays when computing insertion points for barriers. Both of these issues were reported privately and subsequently fixed.
\end{itemize}

\section{Accelerating invariant generation}
\label{sec:performance}

As new rules were integrated into GPUVerify, the responsiveness of the tool diminished.
To illustrate this, we ran a series of experiments over the \LOOPY set using the machine setup from Section~\ref{sec:expsetup}, all with a per-kernel timeout of 10~minutes (all reported times are averages over five runs).
During the first experiment we invoked GPUVerify with user-defined invariants \emph{enabled} but all rules \emph{disabled}.
Then, for each subsequent experiment, we enabled successively more rules according to the order in which we implemented them.
The final experiment therefore enabled all rules at our disposal.
For each run, we recorded the times consumed by GPUVerify to process trivial and non-trivial kernels, including any timeouts.
We split the measurements in this fashion to assess the effect of rule introduction on trivial kernels.

\begin{figure}
\includegraphics[width=\linewidth]{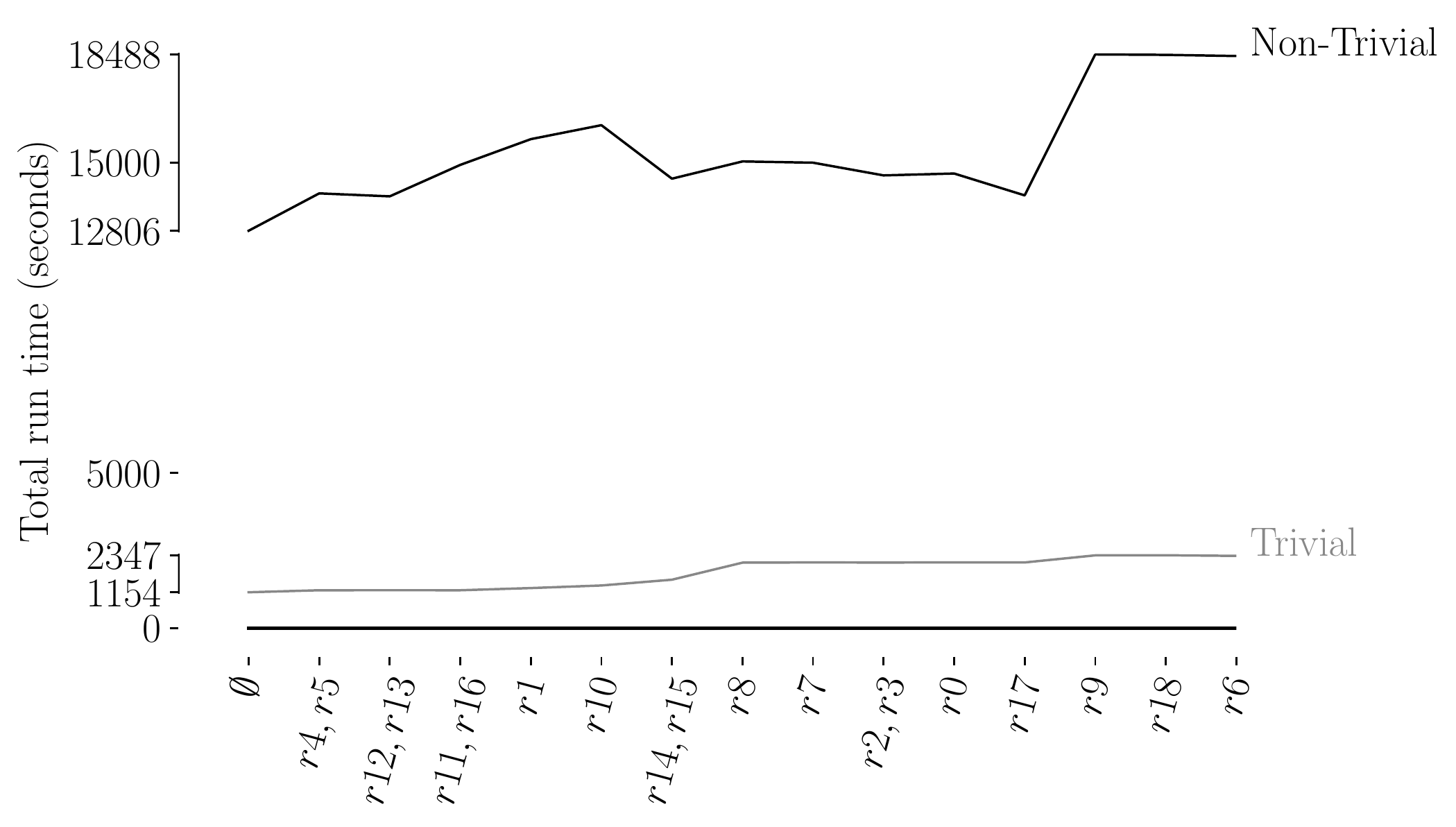}
\caption{The evolution of GPUVerify's performance.}
\label{fig:evolution:performance}
\end{figure}

\begin{figure}
\includegraphics[width=\linewidth]{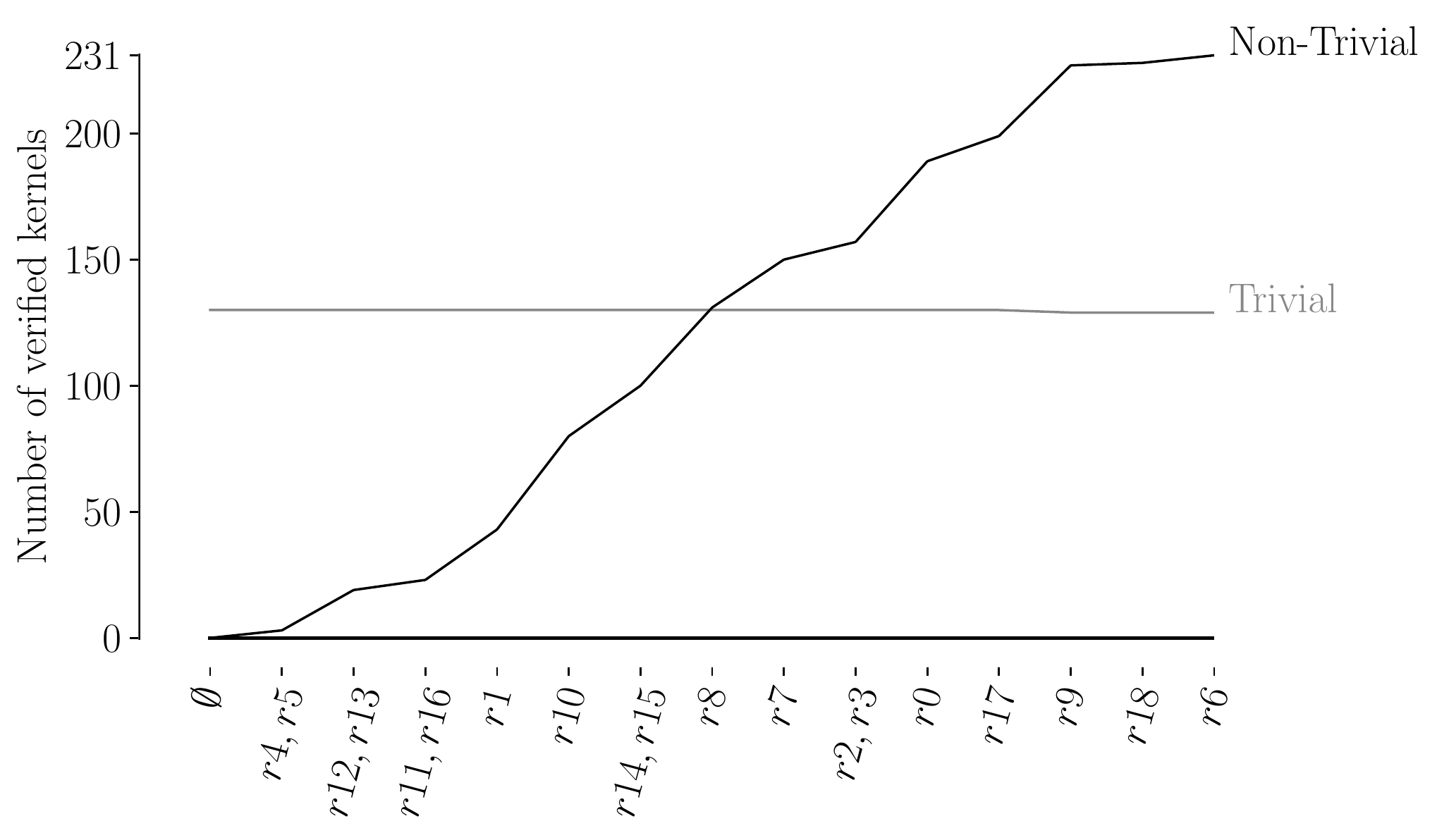}
\caption{The evolution of GPUVerify's verification capabilities.}
\label{fig:evolution:precision}
\end{figure}

The results appear in Figure~\ref{fig:evolution:performance}. The x-axis shows the evolution of the rules, and the two lines plotted show the evolution for the trivial and non-trivial kernels, respectively.  This experiment also allowed us to track the number of kernels GPUVerify was able to verify as the tool evolved. The results of this are shown in Figure~\ref{fig:evolution:precision}. Again, the x-axis shows the evolution of the rules, and the two lines distinguish between there trivial and non-trivial kernels.

Together, Figures~\ref{fig:evolution:performance} and~\ref{fig:evolution:precision} give some insight into the trade-off between precision and performance when increasing the candidate-based invariant generation facilities of a tool.  Figure~\ref{fig:evolution:precision} demonstrates the steady improvement in the number of kernels that GPUVerify could verify,\footnote{Figure~\ref{fig:evolution:precision} also supports the hypothesis from Section~\ref{sec:ruleworth} that rules r4, r8, and r15 were essential when introduced (although they no longer are at present).} but Figure~\ref{fig:evolution:performance} shows an overall increase in analysis time.  The change in analysis time for the non-trivial kernels has limited meaning, since the verification status of these kernels has also changed over time.  However, the verification status for the \emph{trivial} kernels should not have changed---since these kernels already verified without provision of invariants---so the steady increase in analysis time for these kernels is undesirable.

The most important data points in Figure~\ref{fig:evolution:performance} are those for rule r18, showing the total run time with all rules enabled.
We see that the introduction of the rules approximately doubled the time needed to verify the trivial kernels.
The overhead is caused by invariant generation, which GPUVerify must attempt for every kernel, trivial or otherwise.
For non-trivial kernels, performance was hit by a modest $1.4\times$ slowdown.
Merging the results reveals that almost two extra hours were needed to process all kernels in the \LOOPY set (from 233 minutes to 346 minutes) once all rules were integrated.

We hypothesized that Houdini was the cause of the performance reduction.
To validate this hypothesis, we measured the time spent at each GPUVerify stage, i.e.\ at each box in Figure~\ref{fig:gpuverify}, during the final experiment described above.
The breakdown of times is as follows:\footnote{These numbers exclude 17 kernels as they exhausted the timeout.}
\begin{itemize}
  \item 482 seconds in the frontend (including candidate generation),
  \item 7781 seconds in Houdini, and
  \item 2180 seconds in the verification stage.
\end{itemize}
As anticipated, Houdini takes up the bulk (74\%) of GPUVerify's run time.
Motivated by this, we next consider techniques that accelerate candidate-based invariant generation.

\subsection{Refutation engines}

Our idea is to design a number of \emph{under-approximations} of programs that are likely to kill candidates quickly.
Here, a program $T$ under-approximates a program $S$ if the correctness of $S$ implies the correctness of $T$, i.e.\ $T$ can fail in the same or \emph{fewer} ways than $S$.  In this case, we equivalently say that $S$ \emph{over-approximates} $T$.

\begin{figure}
\begin{center}
\includegraphics[width=0.9\columnwidth]{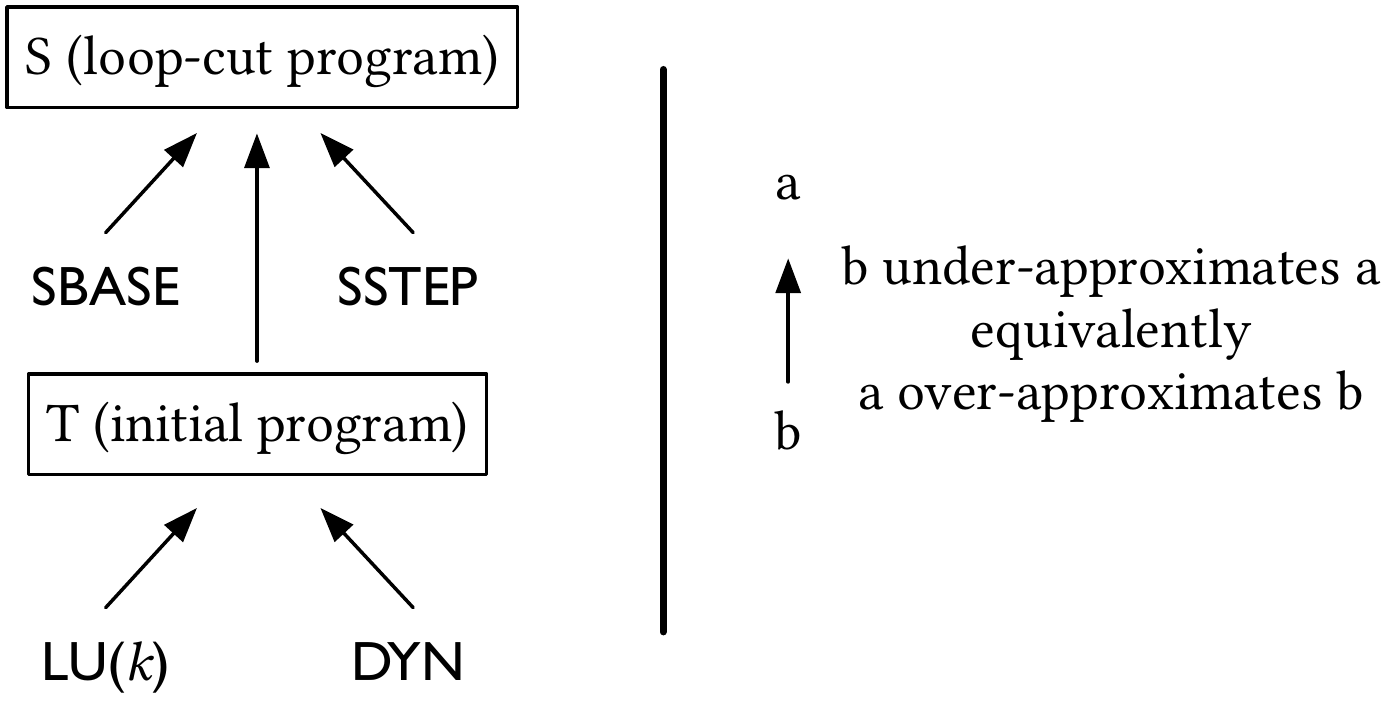}
\end{center}
\caption{%
Given an initial program $T$, Houdini operates on the loop-cut over-approximation $S$.
We propose four refutation engines: variants of the loop-cut program that only check base cases (\SBASE) and step cases (\SSTEP)---both under-approximating $S$---and bounded loop unrolling for depth $k$ (\LU{k}) and dynamic analysis (\DYN)---both under-approximating $T$.
}
\label{fig:engines}
\end{figure}

Houdini employs an over-approximation of the input program---the loop-cut transformation discussed in Section~\ref{sec:loopinv}---to compute an invariant from a set of candidates.
Our idea is to design under-approximations of the loop-cut program that specialize in killing certain types of candidates.
The correctness of our approach rests on a simple observation: any candidate that is shown to be unprovable for an under-approximation of the loop-cut program must also be unprovable for the loop-cut program itself.  We use the term \emph{refutation engine} to refer to an analysis that employs an under-approximation.
With enough time, Houdini will eventually uncover all unprovable candidates, thus a refutation engine is only useful if it finds unprovable candidates ahead of Houdini or it allows Houdini to navigate a faster path through its search space.
We have conceived four refutation engines that we speculated might meet this specification and which we investigate in the remainder of this section (see Figure~\ref{fig:engines} for a summary of the relationships between the engines).

\subsubsection{Splitting loop checking: SBASE and SSTEP}

Recall from Section~\ref{sec:loopinv} that the loop cutting transformation allows us to establish inductive invariants.
As seen in Figure~\ref{fig:loopcut}, the invariant must both hold on loop entry (the base case) and must be maintained by the loop (the step case).
Omitting either of the assertions yields an under-approximation because the resulting program can fail in fewer ways than the original.
This gives us two under-approximations of the loop-cut program $S$: one that only keeps the base cases (\SBASE) and one that only keeps the step cases (\SSTEP).
We can also think of these under-approximations as splitting the program $S$ into two subprograms (c.f.\ Figure~\ref{fig:loopsplit}).
We speculated that refuting candidates in each subprogram separately may be faster than dealing with the program as a whole:
although we might expect the sum of the times taken to prove the base and step cases separately to be similar to the time associated with proving them in combination, if we have a base case (or step case) that does \emph{not} hold due to an unprovable candidate invariant, we might be able to establish this more quickly by considering the base case (or step case) in isolation, leading to faster refutation of the offending candidate.

\begin{figure}
\begin{minipage}[p]{0.49\columnwidth}
\begin{lstlisting}
// base case
assert $\phi$;
havoc modset(B);
assume $\phi$;
if (c) {
  B;
  // step case omitted
  assume false;
}
\end{lstlisting}
\end{minipage}
\hfill
\begin{minipage}[p]{0.49\columnwidth}
\begin{lstlisting}
// base case omitted
havoc modset(B);
assume $\phi$;
if (c) {
  B;
  // step case
  assert $\phi$;
  assume false;
}
\end{lstlisting}
\end{minipage}
\caption{Splitting the loop checking for the loop-cut program $S$ yields two under-approximations, \SBASE (left) and \SSTEP (right).}
\label{fig:loopsplit}
\end{figure}

\subsubsection{Bounded loop unrolling, LU($k$)}

Bounded loop unrolling a program for a given depth $k$ yields a loop-free program where only the first $k$ iterations of each loop are considered.  When applied to a program $T$ this yields an under-approximation of $T$.  The method is commonly employed by bounded model checking tools such as CBMC~\cite{CBMC}.
Figure~\ref{fig:loopunroll} shows the transformation of a loop after unrolling for depth $k=2$.
The loop-free fragment models $k$ iterations of the loop.
The resulting program is an under-approximation because it does not consider behaviors that require further loop iterations.
The \code{assume false} statement implies that any execution that would continue past $k$ iterations is infeasible and will not be considered~\cite{WeakestPrecondition05}.

Despite encoding only a subset of the original program's behavior, loop unrolling leads to a syntactically larger program that, when converted to an SMT formula, may place a high burden on the underlying SMT solver.  This is especially problematic in the case of nested loops, where unwinding an outer loop $k$ times creates $k$ copies of all inner loops, which must then be unwound in turn.  For this reason, in our experiments we consider only the \LU{1} configuration, where loops are unwound up to depth one.  A key difference between \SBASE and \LU{1} is that all program loops are abstracted when \SBASE is employed, while \LU{1} uses no abstraction.  This means e.g.\ that if two loops appear in sequence, and the first loop has an iteration count of at least two, no candidates of the second loop can be eliminated when \LU{1} is used since every program path that reaches the second loop involves two or more loop iterations.  In contrast, \SBASE considers program paths that abstract the first loop, reaching the head of the second loop directly.

\begin{figure}
\begin{minipage}[p]{0.35\columnwidth}
\begin{lstlisting}
while (c)
  invariant $\phi$ {
  B;
}
\end{lstlisting}
\end{minipage}
\hfill
\begin{minipage}[p]{0.45\columnwidth}
\begin{lstlisting}
if (c) {
  assert $\phi$;
  B;
  if (c) {
    assert $\phi$;
    B;
    if (c) {
      assume false;
    }
  }
}
\end{lstlisting}
\end{minipage}
\caption{%
Bounded loop unrolling of the loop on the left for depth $k=2$ yields the loop-free program on the right.
}
\label{fig:loopunroll}
\end{figure}

\subsubsection{Dynamic analysis, DYN}
\label{sec:engines:dynamic}

Executing a program $T$ is a classic under-approximating analysis which, unlike our other refutation engines, is not dependent on a SMT solver.
Instead, the statements of $T$ are simply interpreted.
To enable execution, we implemented an interpreter for \emph{Boogie}---the intermediate verification language into which GPUVerify translates kernels and in which it expresses candidate invariants.\footnote{Boogaloo~\cite{Boogaloo} and Symbooglix~\cite{Symbooglix} also support Boogie interpretation, but are generic and do not exploit knowledge specific to GPU kernels, as we do.}

Our dynamic analyzer executes each kernel multiple times.
Before each invocation, values for formal parameters and thread and block identifiers (i.e.\ \code{threadIdx} and \code{blockIdx}) are chosen that satisfy the preconditions of the kernel (c.f.\ Section~\ref{sec:methodology:params}).
Re-invocation halts once a selected coverage criterion---basic block coverage---is met or a specific number of launches has been reached.
For many kernels we find that a \emph{single} execution suffices to achieve full basic block coverage, because GPU code is rarely control dependent on formal parameters or thread variables.
This means we can simply choose random values and can ignore sophisticated test-generation techniques, which is clearly not applicable to other domains.
In spite of this simplicity, dynamic analysis may still be slow, for two reasons.

First, much execution time may be spent in loops with large bounds without refuting many candidates.
Typically, this is due to dynamic analysis rejecting a candidate on the \emph{first} loop iteration, or not at all.
Hence, iterating through loops does not serve our aim of accelerating invariant generation.
Our pragmatic solution is to bound the number of loop iterations summed across all loops.
The downside is that a single loop may hog the execution, preventing analysis of candidates in other loops.
This drawback is more severe if there are candidates in loops after the cut-off point that are easily disproved through dynamic analysis but difficult to reject through an SMT-based refutation engine.

Second, candidates involving reads from and writes to arrays should be evaluated for all array indices discovered during the dynamic analysis.
For instance, suppose we have arrays $A$ and $B$ and a candidate asserting that all accesses into $A$ and into $B$ are distinct.
Then, we must evaluate this candidate with respect to all tuples $(a, b)$, where $a$ and $b$ are \emph{observed} array indices of $A$ and $B$, respectively.
Checking all tuples, however, is generally not feasible as the number grows exponentially in the length of the tuple.
Instead, we select a constant number of random tuples, using the rationale that a candidate is likely true if it holds for this restricted subset.
An obvious disadvantage is that the random selection may miss an instance that falsifies the property.

A risk associated with employing dynamic analysis is that the semantics of the dynamic analyzer might diverge (unintentionally) from the semantics of Boogie.  This could lead to the refutation of candidates that Houdini would in fact be able to prove, or to the failure to refute candidates that a correct dynamic analysis would identify as false.  While we took care to implement an accurate analysis, we note that this risk cannot compromise the soundness of verification.  In the first case, where dynamic analysis refutes a provable candidate, the price may be that verification of the kernel subsequently fails, due to an insufficiently strong invariant being inferred.  In the latter case, where dynamic analysis fails to refute a candidate, the candidate is guaranteed to be refuted eventually by Houdini, so the price is merely that the performance benefit of using dynamic analysis may not be realized.

\subsection{Evaluation of refutation engines}

We conducted several experiments addressing the following questions:
\begin{itemize}
  \item Is a refutation engine able to reject candidates?
  \item Do refutation engines complement each other?
  \item Is invariant generation accelerated by a refutation engine?
  \item Does launching multiple refutation engines in parallel yield discernible gains?
\end{itemize}

\subsubsection{Experimental setup}

For these experiments we were interested in measuring the performance of GPUVerify using various invariant generation strategies.  As the issue of performance fluctuation across platforms is well-known, we performed the experiments across two machines running different operating systems; this to reduce measurement bias~\cite{MeasurementBias}:
\begin{itemize}
  \item a \emph{Windows} machine with a 2.4GHz 4-core Intel Xeon E5-2609 processor and 16GB of RAM, running Windows 7 (64 bit) and using CVC4 v1.5-prerelease, Clang/LLVM v3.6.2, and Common Language Runtime v4.0.30319; and

  \item a \emph{Ubuntu} machine also with a 2.4GHz 4-core Intel Xeon E5-2609 processor and 16GB of RAM, running Ubuntu 14.04 and using CVC4 v1.5-prerelease, Clang/LLVM v3.6.2, and Mono v3.4.0 (this is the same machine as the one used in the precision experiments).
\end{itemize}

The four refutation engines considered were \SBASE, \SSTEP, \LU{1}, and \DYN.
The dynamic analyzer had the following settings (c.f.\ Section~\ref{sec:engines:dynamic}), which we obtained via exploratory manual tuning during development of the analyzer: it quit as soon as 100\% basic block coverage was met or 5 executions completed; it terminated a single execution if 1,000 loop iterations were processed; a candidate referring to tuple of array indices was evaluated with respect to 5 distinct, randomly chosen tuples of observed values (or fewer, if fewer than 5 distinct tuples had been observed).
All these experiments included our user-defined invariants (the PolyBench/C suite also included the compiler-generated invariants discussed in Section~\ref{sec:candidategenerationprocess}), and used a timeout of 10~minutes.
All reported times are averages over five runs.

\subsubsection{Experiment: refutation engine power}\label{sec:throughput}

The first hypothesis we wished to validate was whether every refutation engine could reject candidates at least as fast as Houdini.
To this end, we ran each refutation engine in isolation and measured both the time consumed and the number of candidates refuted.
We present the results in Table~\ref{tab:enginethroughput}, showing numbers for Houdini (denoted by \HOU) for comparative purposes.

\begin{table*}
\caption{Refutation engine performance and throughput.}
\begin{center}
\begin{tabular}{l|rrr|rrr}
 & \multicolumn{3}{c|}{\textbf{Windows}} &  \multicolumn{3}{c}{\textbf{Ubuntu}}\\
Engine & Refutations & Total time & Throughput          &  Refutations & Total time & Throughput\\
            &                     & (sec)        & (refutations/sec)  &                     & (sec)        & (refutations/sec)\\
\toprule
\HOU   & 5,703 & 17,805 & 0.32 & 5,615 & 15,544 & 0.36 \\ 
\SBASE & 3,692 & 5,053  & 0.73 & 3,692 & 4,991  & 0.74 \\ 
\SSTEP & 3,421 & 15,125 & 0.23 & 3,430 & 14,664 & 0.23 \\ 
\LU{1} & 3,712 & 10,096 & 0.37 & 3,754 & 9,541  & 0.39 \\ 
\DYN   & 2,367 & 811   & 2.92 & 2,301 & 2,828  & 0.81 \\ 
\end{tabular}
\end{center}
\label{tab:enginethroughput}
\end{table*}

The yardstick in this experiment is throughput: the number of refutations per second.
We see that \DYN is extremely effective on Windows, with a throughput that is four times that of the next-best performing refutation engine, but much less so on Ubuntu, where the difference to the next-best performing engine is marginal.
The throughputs of the other refutation engines appear mostly insensitive to the machine setup; we attribute the discrepancy in throughput for \DYN to differences in the Common Language Runtime implementation.
\SBASE has a high throughput on both machines and is much more effective than \SSTEP, suggesting that it is easier for the SMT solver to reason about base case candidates.
\LU{1} has a moderate throughput, but kills the most candidates among the refutation engines.

The results indicate that \DYN and \SBASE show promise for acceleration of candidate refutation, while \LU{1} has only marginally higher throughput compared with \HOU.  The throughput of \SSTEP is poor.  Anecdotally, our experience working with SMT-based program verifiers is that SMT solvers tend to spend more time reasoning about the step case associated with a loop compared with the base case.  We thus hypothesize that the \SBASE engine achieves high throughput by rapidly refuting all invariants that can be eliminated via base-case-only reasoning, avoiding the hard work associated with the step case.  In contrast, \SSTEP undertakes the work that is hard for \HOU---the step case---but unlike \HOU does not provide throughput by eliminating candidates that can only be refuted by base case reasoning.

\subsubsection{Experiment: complementary power}

Our expectation was that refutation engines would each specialize in killing candidates generated by different rules, and that the refutation engines would therefore complement each other.
To test this, we recorded the set of candidates rejected by a refutation engine across the \emph{whole} \LOOPY set and then calculated the Jaccard index~\cite{Jaccard} between the sets for every pair of refutation engines.
The Jaccard index numerically evaluates the similarity among sets:
\[
    J(A, B) = \frac{\lvert A \cap B \rvert}{\lvert A \cup B \rvert} \, .
\]
For non-empty sets $A$ and $B$, $J(A, B) = 1$ if the sets are identical, and $J(A, B) = 0$ if the sets share no common elements.
The higher the Jaccard index, the more related the sets are.
In our case, when comparing the sets of candidates killed by two distinct refutation engines, a low Jaccard index indicates that the two engines are complementary in their refutation power.

\begin{table}
\caption{Refutation engine similarity in terms of refuted candidates; a low Jaccard index indicates that two engines are complementary in their refutation power}
\begin{center}
\begin{tabular}{ll|r|r}
   & & \multicolumn{1}{c|}{\textbf{Windows}} &  \multicolumn{1}{c}{\textbf{Ubuntu}}\\
\multicolumn{2}{c|}{Refutation engine pair} & Jaccard index & Jaccard Index\\
\toprule
\DYN & \SBASE & 0.12 & 0.17 \\
\DYN & \SSTEP & 0.23 & 0.25 \\
\DYN & \LU{1} & 0.22  & 0.25 \\
\SBASE & \SSTEP & 0.33 & 0.39 \\
\SBASE & \LU{1} & 0.63 & 0.70 \\
\SSTEP & \LU{1} & 0.49 & 0.53\\
\end{tabular}
\end{center}
\label{tab:jaccard}
\end{table}

Table~\ref{tab:jaccard} gives the Jaccard indices.
We observe that \DYN complements every SMT-based refutation engine, especially \SBASE.
Given that \DYN and \SBASE were also the best performing engines in the throughput experiment, we hypothesized that these refutation engines together would be able to accelerate invariant discovery (we further investigate this hypothesis in the next section).
Finally, the higher Jaccard index for \SBASE and \LU{1} suggests that these engines refute similar candidates, while the lower Jaccard index for \SBASE and \SSTEP indicates, as expected, that these engines target different candidates.

Note that the computed Jaccard indices differ between our experimental machines because of our use of a timeout.  With enough time, our refutation engines would compute the same results on both platforms.  However, as our aim is to accelerate invariant generation, the extent to which a refutation engine can refute candidates is of limited interest if the time required to do so is excessive.

\subsubsection{Experiment: overall performance impact}

A drawback of evaluating a refutation engine in terms of throughput is that this disregards the difficulty of refuting the remaining unprovable candidates.
If a refutation engine merely quashes easily disproved candidates, then Houdini must still do the heavy lifting.
The experiment described in this section  therefore assesses whether the proposed refutation engines, or a combination thereof, actually help Houdini to reach a fixpoint faster.

We compared the time to return an invariant for a kernel across various refutation engine configurations.
Our baseline configuration was Houdini \emph{in isolation}, which is consistent with the current state of the art.
We set up a number of \emph{sequential} configurations whereby Houdini ran after a refutation engine had terminated; these configurations are denoted $\textbf{R};\HOU$ where $\textbf{R} \in \{\DYN, \SBASE, \SSTEP, \LU{1}\}$.
We also considered a single \emph{parallel} configuration whereby \DYN and \SBASE were launched alongside Houdini; this configuration is denoted $\DYN\|\SBASE\|\HOU$.
We selected \DYN and \SBASE for our parallel configuration because of their high throughputs and complementary nature, as observed in our previous experiments.

In the parallel configuration, there is a shared pool of refutations that Houdini reads on each iteration.
The exchange of rejected candidates is therefore \emph{asynchronous}.
An asynchronous exchange is allowed for two reasons:
\begin{enumerate}
  \item Houdini guarantees that the number of candidates decreases in a strictly monotonic fashion~\cite{HoudiniMonotonic}, and
  \item every candidate killed by a refutation engine may be trusted (because the engine employs an under-approximating analysis).
\end{enumerate}
Note that the completion time of the parallel configuration is measured as the time for Houdini to terminate; at that point the refutation engines may still be running, but an invariant has been computed.

\begin{figure*}
   \centering
   \subfloat[Total run times]{
   \label{fig:speedups:windows:a}
    \includegraphics[width=0.475\textwidth]{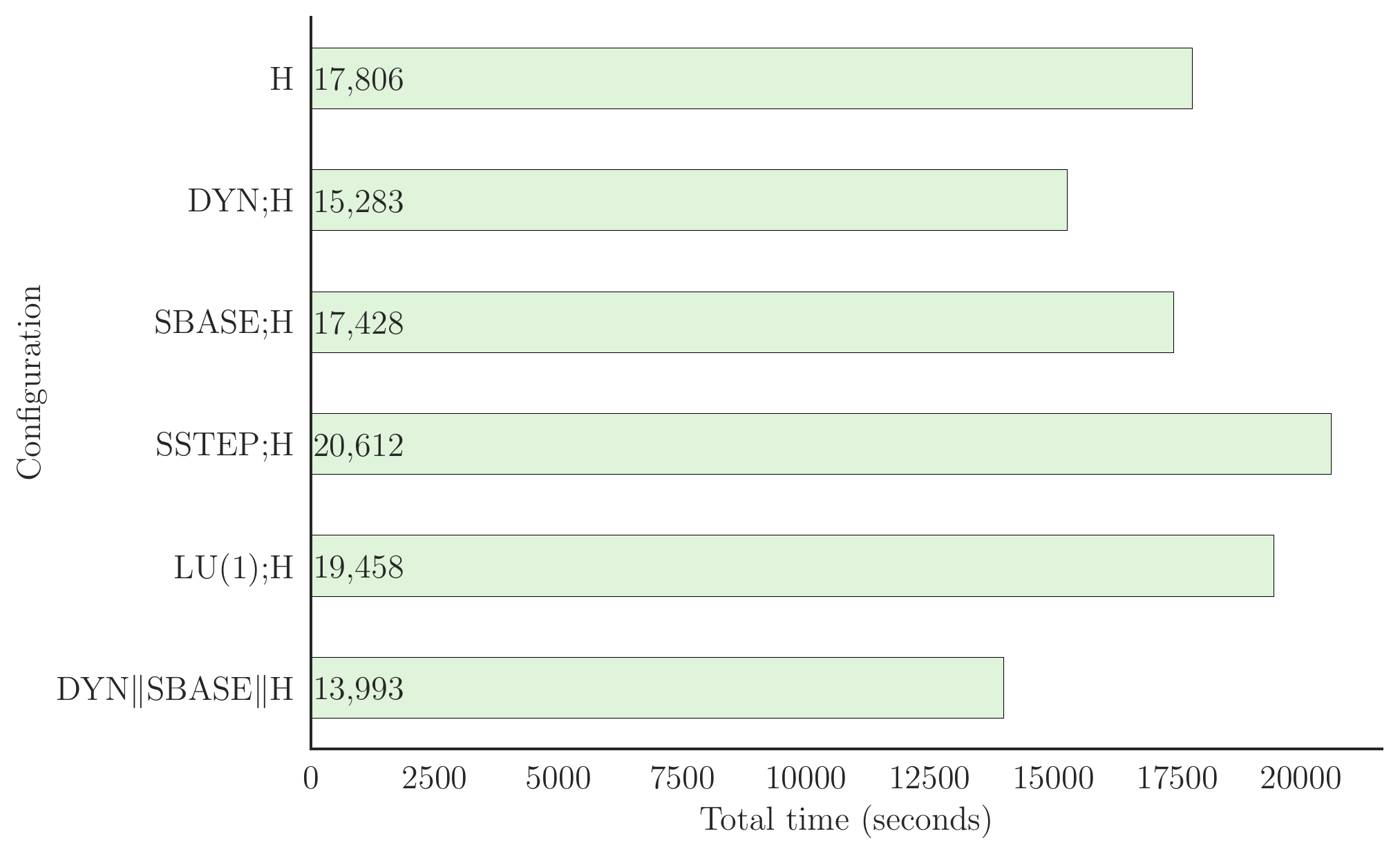}
    }
    \subfloat[\DYN;\HOU]{
   \label{fig:speedups:windows:b}
    \includegraphics[width=0.475\textwidth]{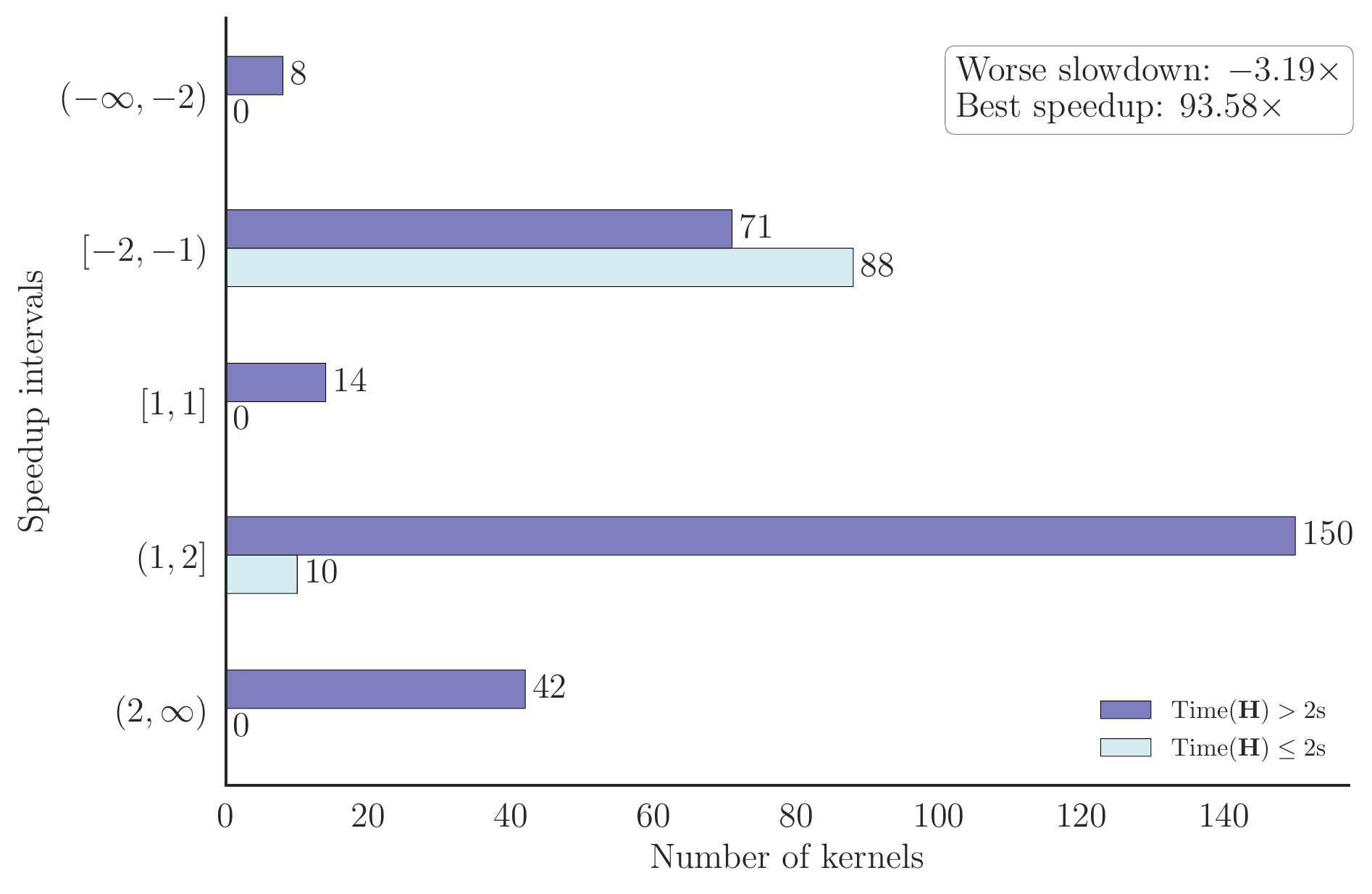}
    }
    \hfill
    \subfloat[\SBASE;\HOU ]{
    \label{fig:speedups:windows:c}
    \includegraphics[width=0.475\textwidth]{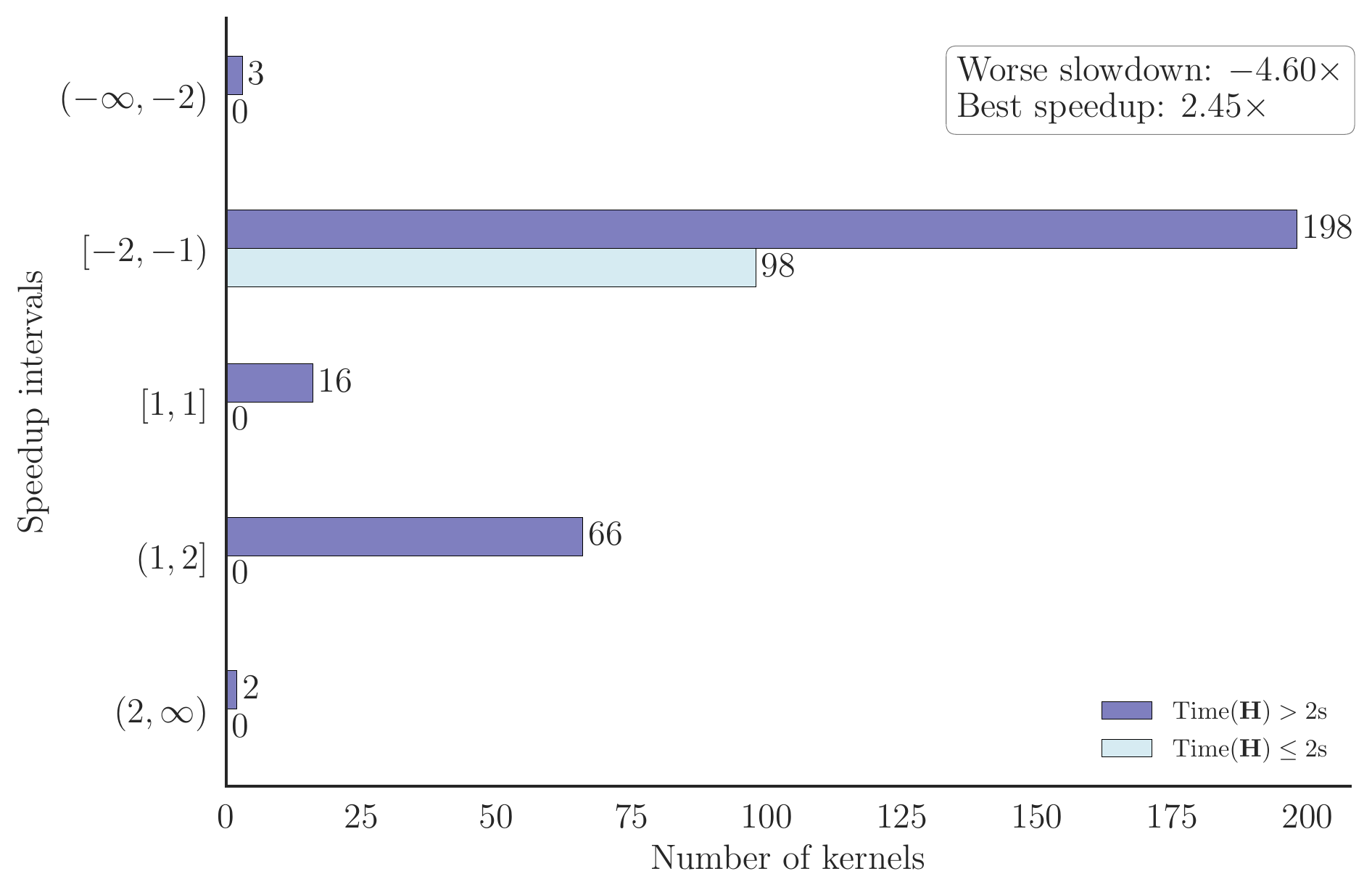}
    }
    \subfloat[\SSTEP;\HOU ]{
   \label{fig:speedups:windows:d}
    \includegraphics[width=0.475\textwidth]{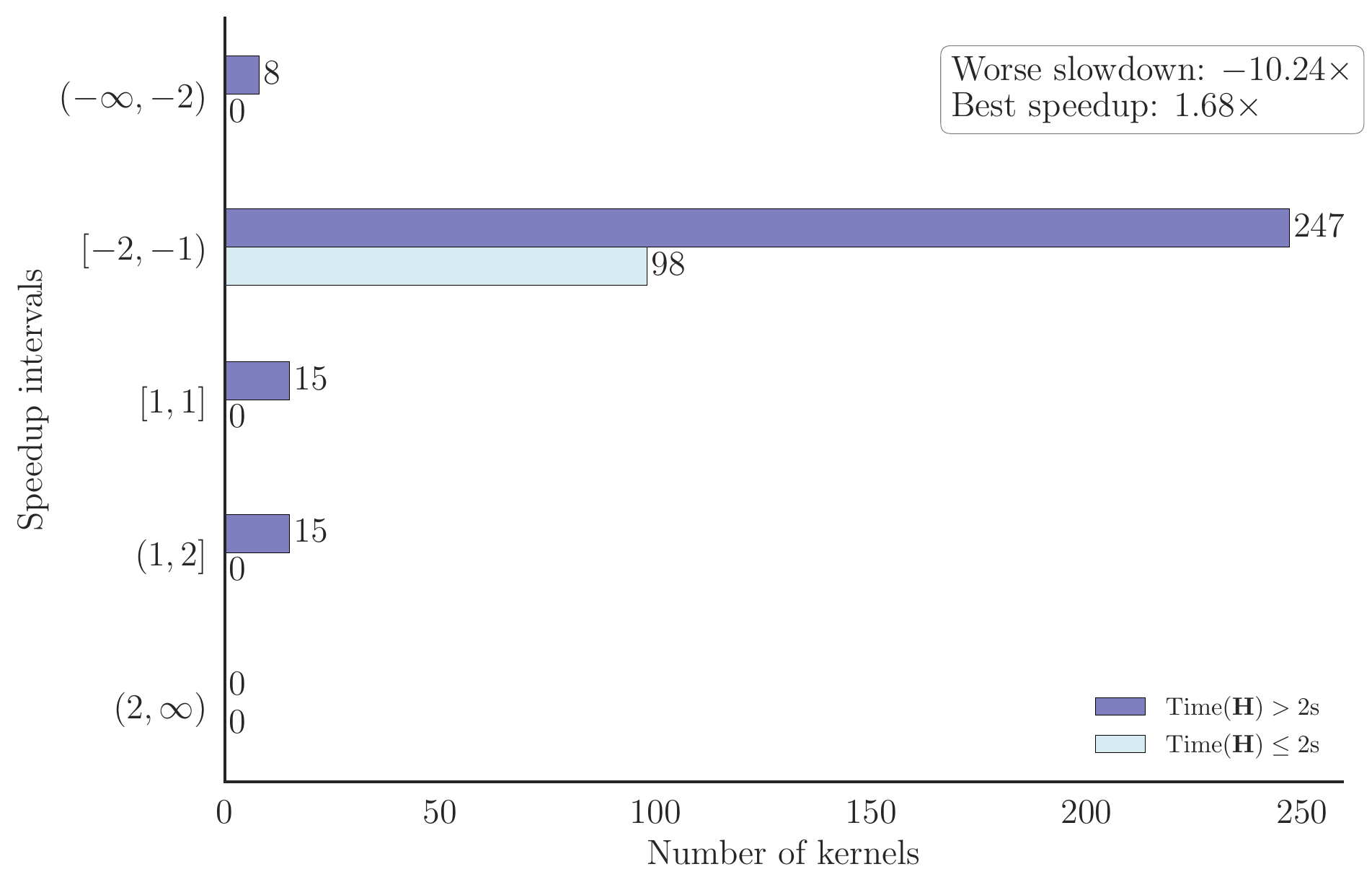}
    }
    \hfill
    \subfloat[\LU{1};\HOU ]{
    \label{fig:speedups:windows:e}
    \includegraphics[width=0.475\textwidth]{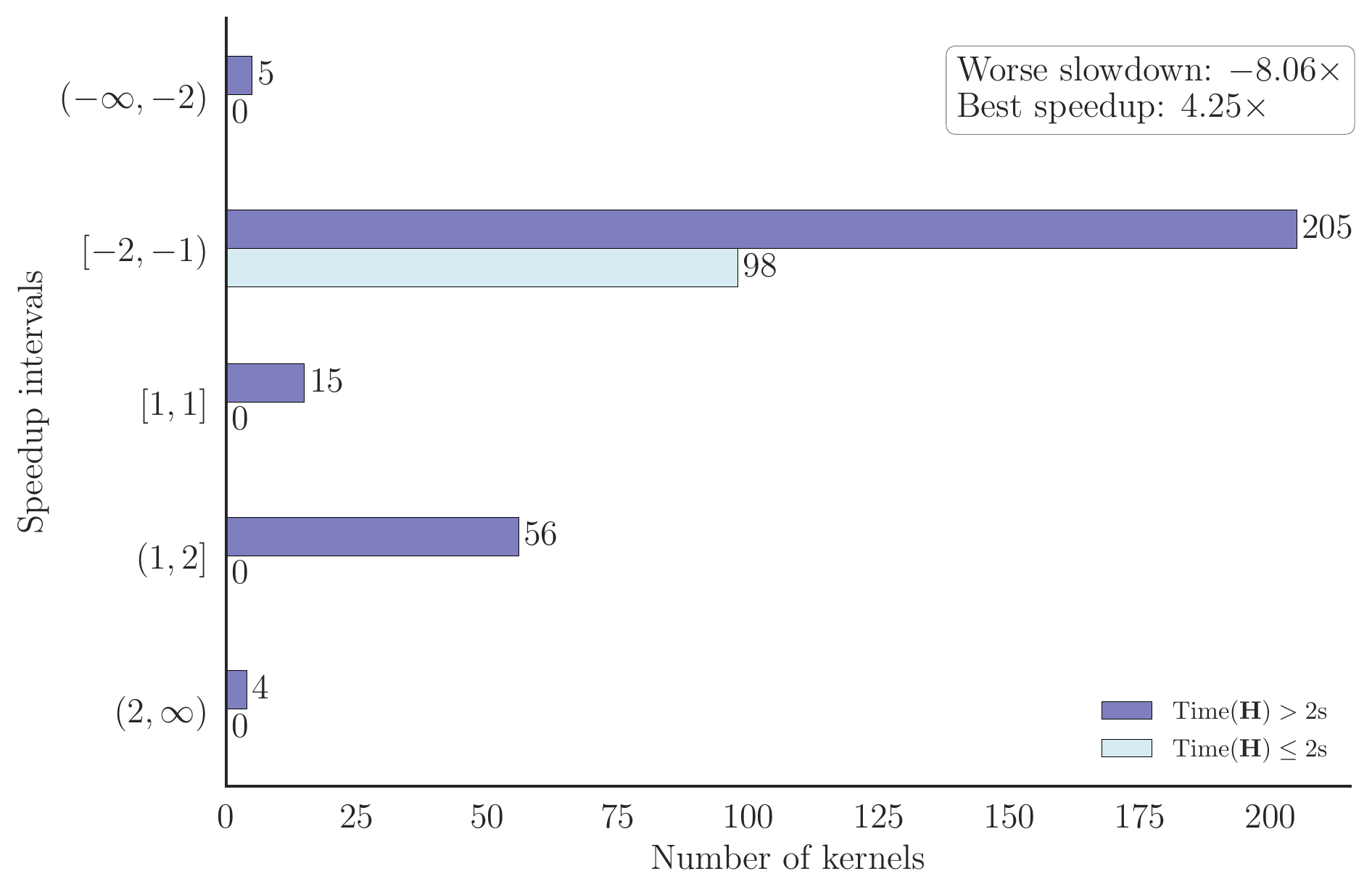}
    }
    \subfloat[$\DYN \| \SBASE \| \HOU$]{
    \label{fig:speedups:windows:f}
    \includegraphics[width=0.475\textwidth]{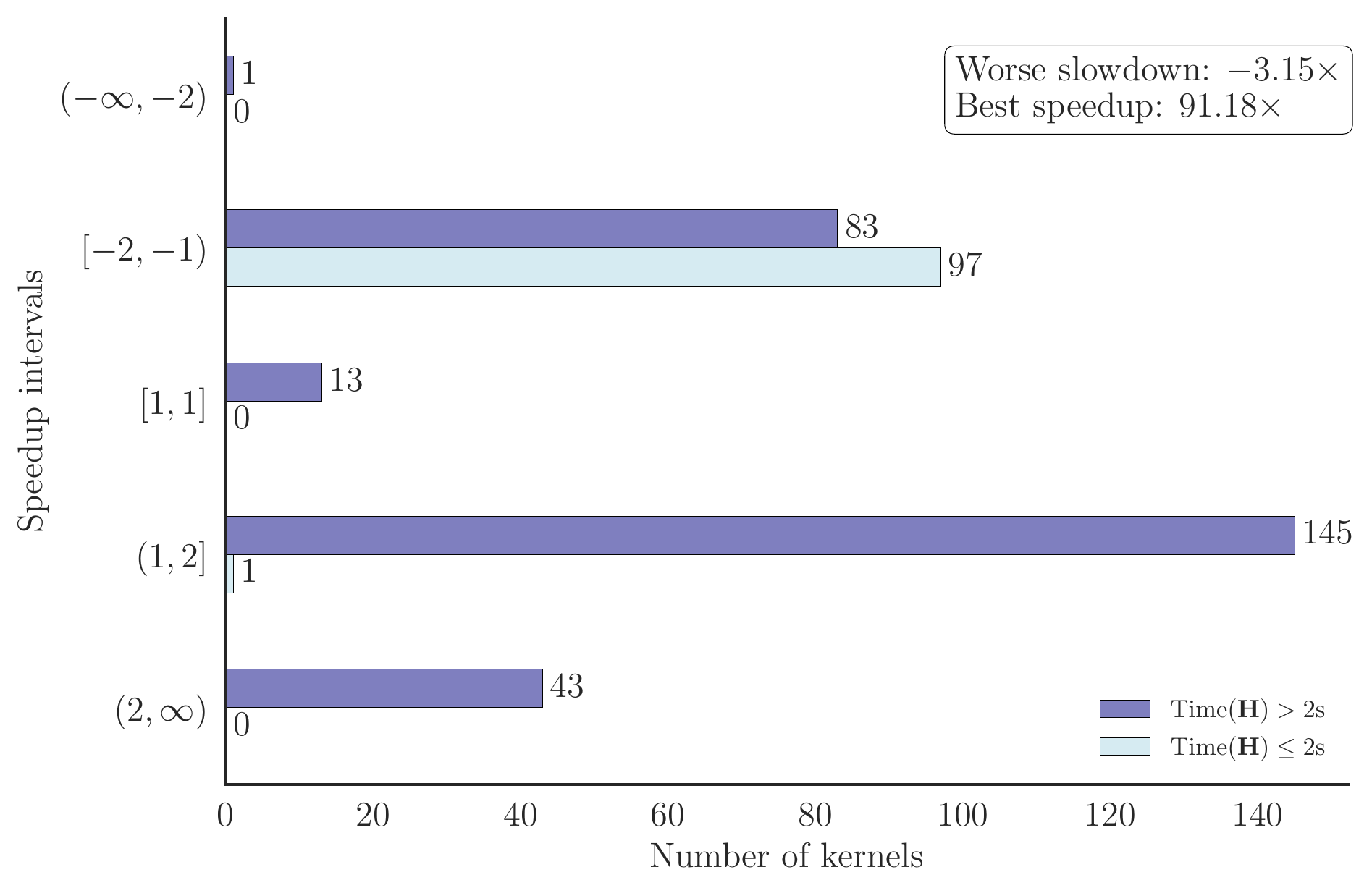}
    }
    \caption{Overall performance impact on the Windows machine, organized by noteworthy slowdowns, $(-\infty,-2)$; modest slowdowns, $[-2,-1)$; break-evens, $[1,1]$; modest speedups, $(1,2]$; and noteworthy speedups, $(2,\infty)$.}
    \label{fig:speedups:windows}
\end{figure*}

\begin{figure*}
   \centering
   \subfloat[Total run times]{
   \label{fig:speedups:ubuntu:a}
    \includegraphics[width=0.475\textwidth]{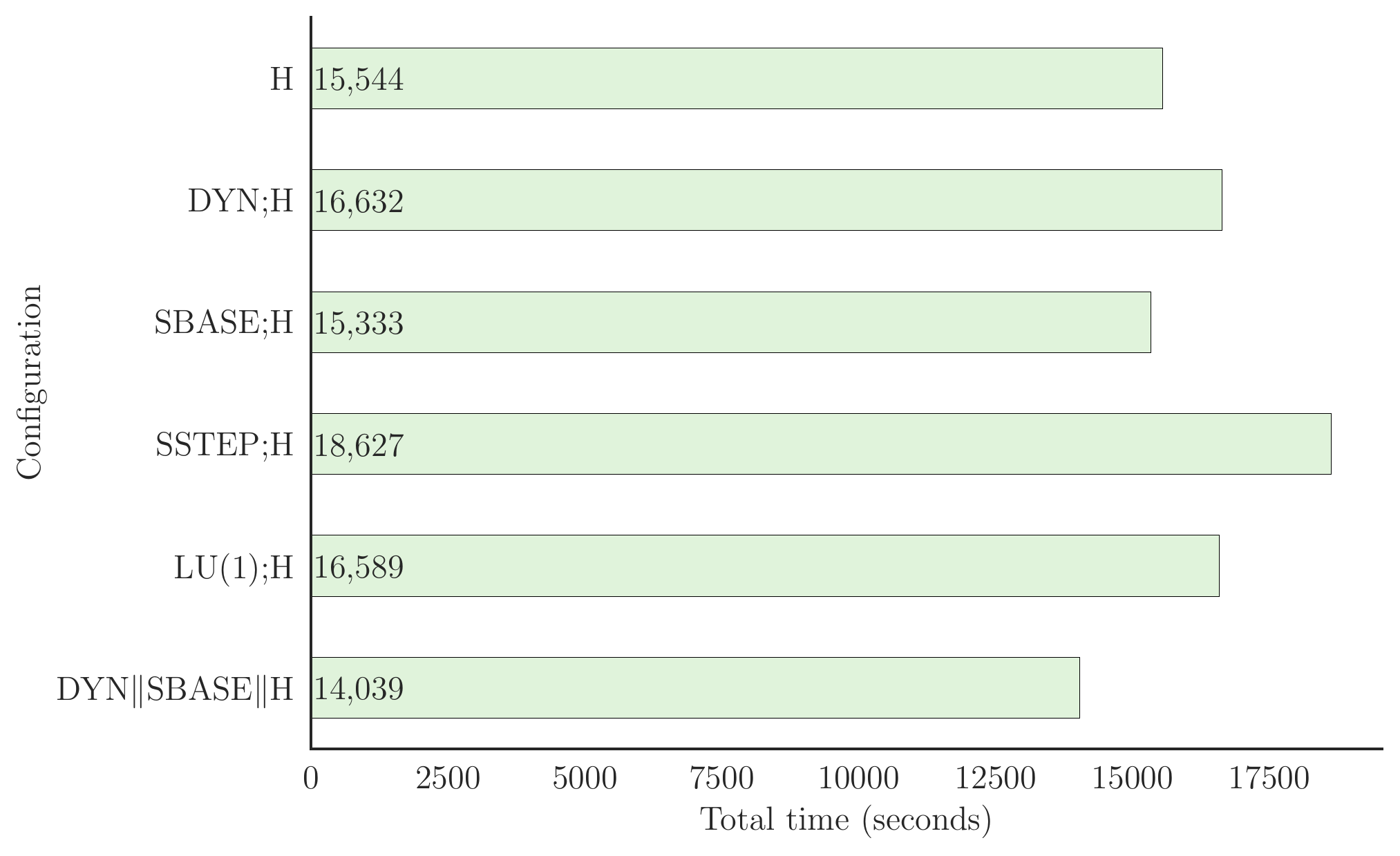}
    }
    \subfloat[\DYN;\HOU]{
    \label{fig:speedups:ubuntu:b}
    \includegraphics[width=0.475\textwidth]{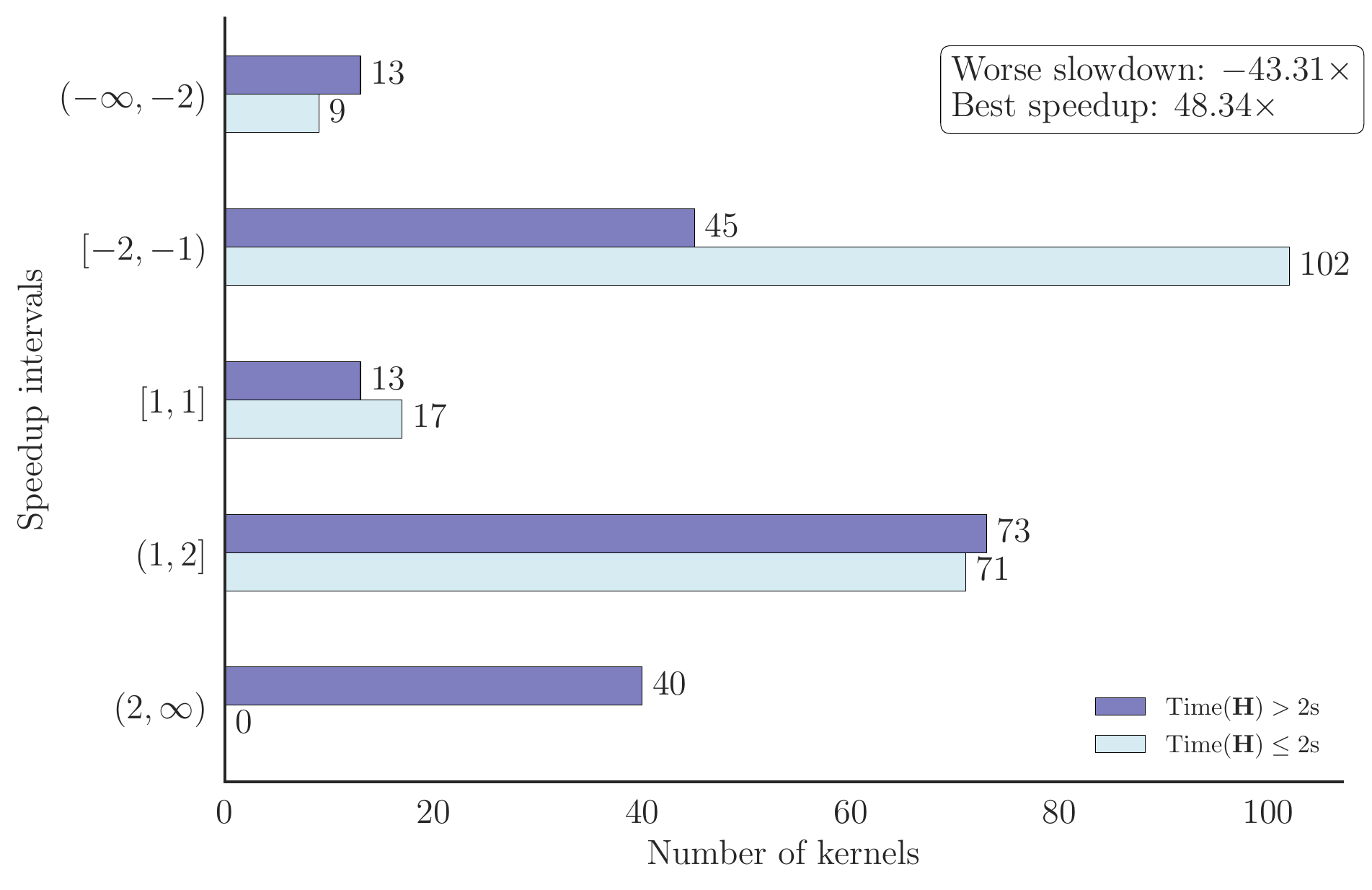}
    }
    \hfill
    \subfloat[\SBASE;\HOU ]{
    \label{fig:speedups:ubuntu:c}
    \includegraphics[width=0.475\textwidth]{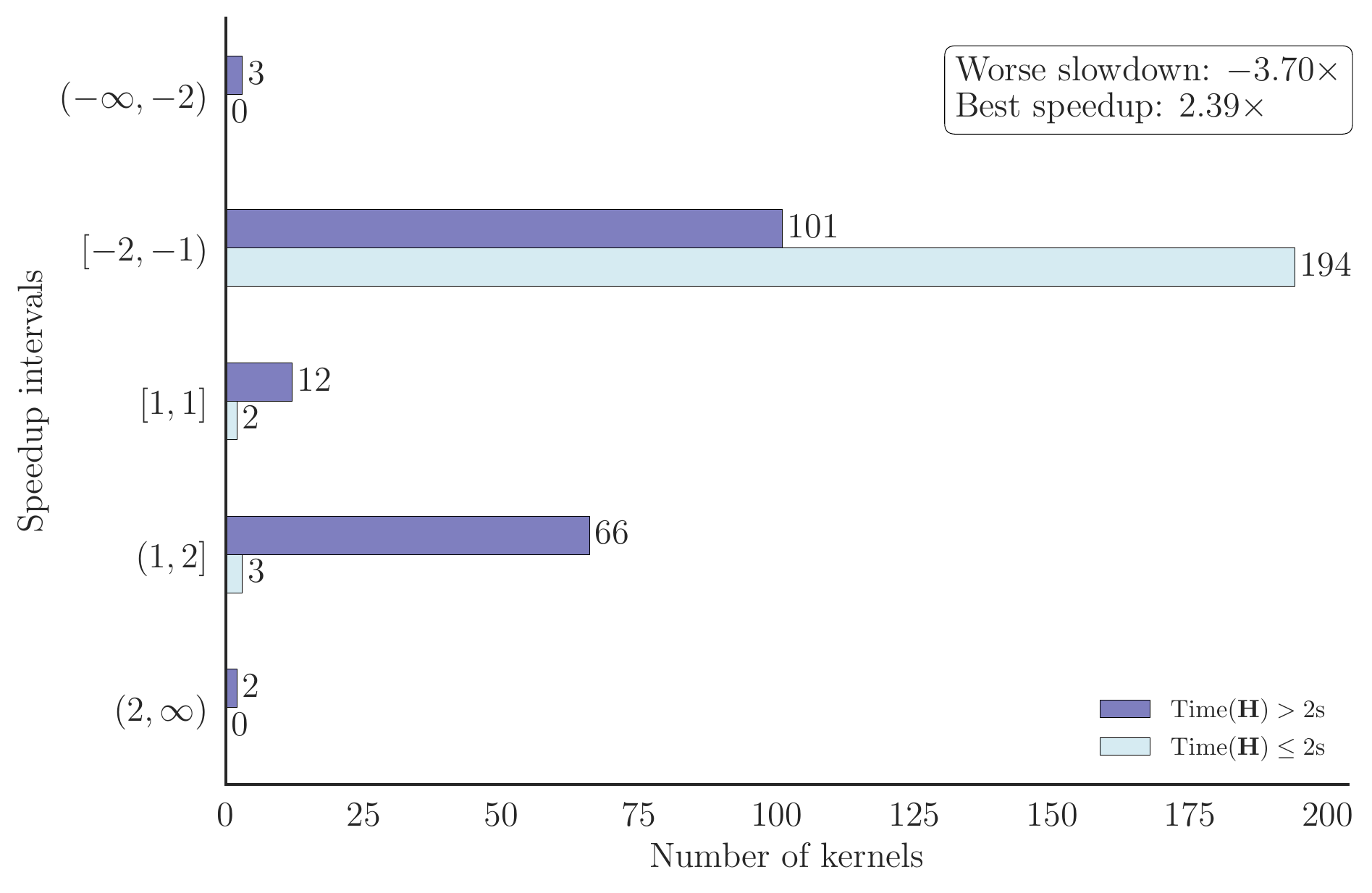}
    }
    \subfloat[\SSTEP;\HOU ]{
    \label{fig:speedups:ubuntu:d}
    \includegraphics[width=0.475\textwidth]{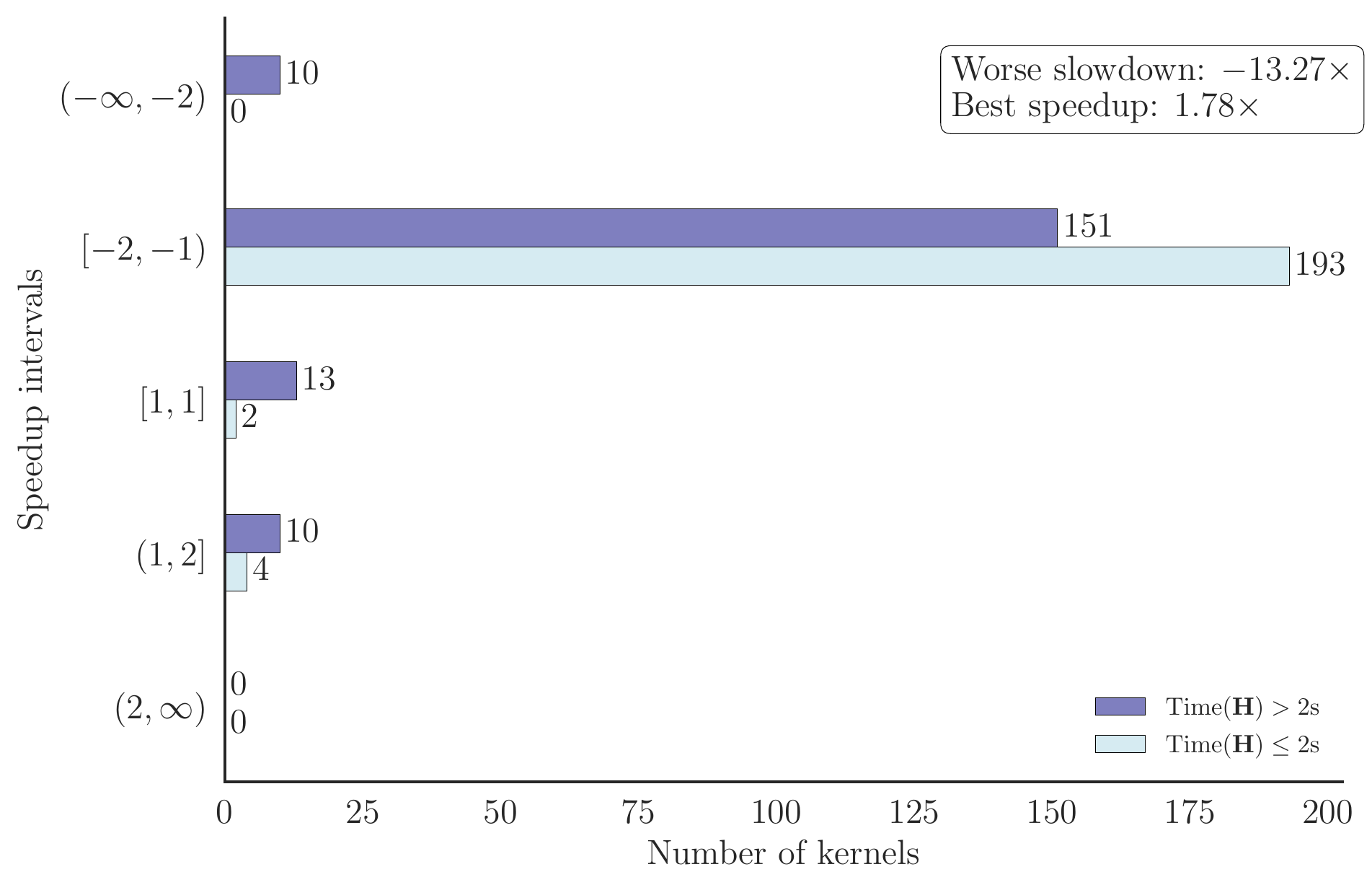}
    }
    \hfill
    \subfloat[\LU{1};\HOU ]{
    \label{fig:speedups:ubuntu:e}
    \includegraphics[width=0.475\textwidth]{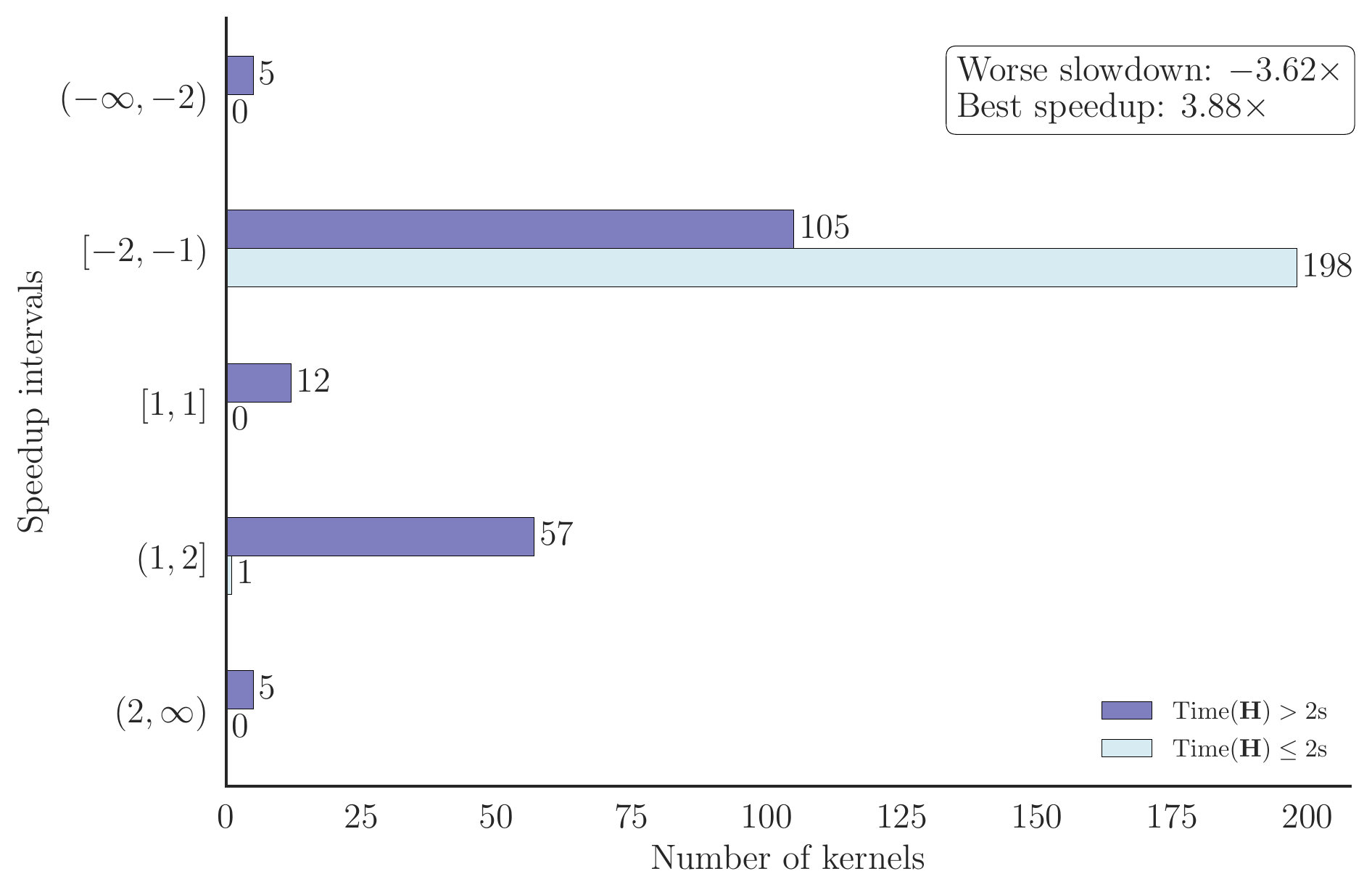}
    }
    \subfloat[$\DYN \| \SBASE \| \HOU$]{
    \label{fig:speedups:ubuntu:f}
    \includegraphics[width=0.475\textwidth]{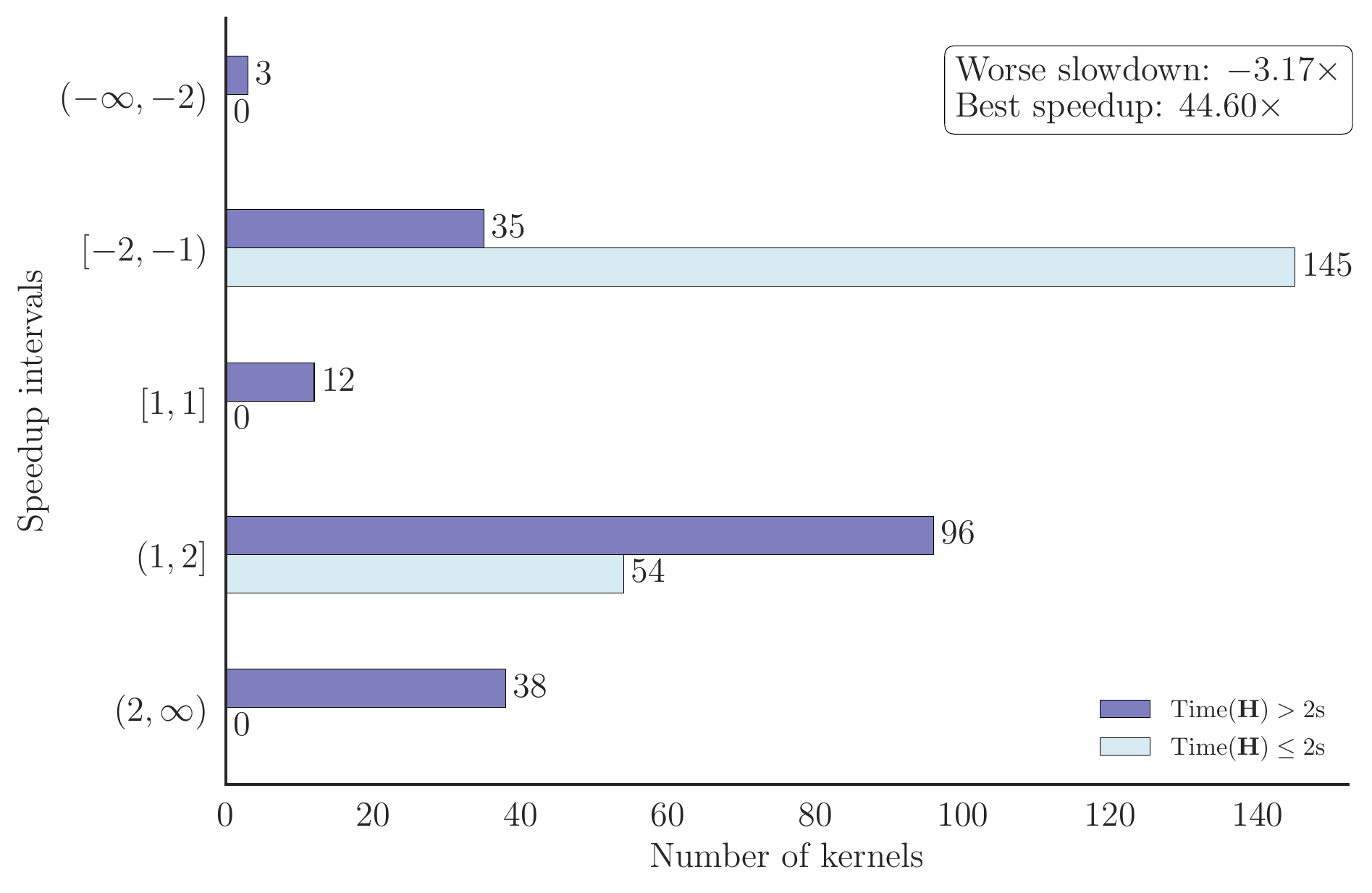}
    }
    \caption{Overall performance impact on the Ubuntu machine, organized by noteworthy slowdowns, $(-\infty,-2)$; modest slowdowns, $[-2,-1)$; break-evens, $[1,1]$; modest speedups, $(1,2]$; and noteworthy speedups, $(2,\infty)$.}
    \label{fig:speedups:ubuntu}
\end{figure*}

Figures~\ref{fig:speedups:windows} and~\ref{fig:speedups:ubuntu} present the results for the Windows and Ubuntu machine, respectively.
There are two types of bar charts.
The first (Figures~\ref{fig:speedups:windows:a} and~\ref{fig:speedups:ubuntu:a}) provides a bird's-eye view of performance, showing the total times to process all kernels in the \LOOPY set for each configuration.
The second (Figures~\ref{fig:speedups:windows:b}--\ref{fig:speedups:windows:f} and~\ref{fig:speedups:ubuntu:b}--\ref{fig:speedups:ubuntu:f}) narrows the focus to a specific configuration, grouping per-kernel performance comparisons into five intervals: $(-\infty,-2)$ (noteworthy slowdowns), $[-2,-1)$ (modest slowdowns), $[1,1]$ (break-evens), $(1,2]$ (modest speedups), and $(2,\infty)$ (noteworthy speedups).
Each of these intervals is divided into two categories depending on whether we deem invariant refutation for a kernel to be inherently \emph{fast} ($\leq 2$ seconds for the baseline configuration to finish) or inherently \emph{slow} ($> 2$ seconds for the baseline configuration to finish).
We split the intervals to be able to evaluate whether the speedups and slowdowns of a configuration are actually noticeable to a user.
Any improvement in speed is likely to be more noticeable when invariant refutation is slow while, conversely, any performance deterioration is likely more noticeable when invariant refutation is fast.
The threshold of two seconds is simply based on our experience to date.
The break-evens ($[1,1]$) indicate that any change in analysis time is absorbed by floating-point round-off error.  In most instances these are due to kernels for which analysis times out with both Houdini and the configuration under consideration, while some cases are accounted for by kernels for which the analysis time is very fast, so that performance differences are likely below the granularity of the system clock used for time measurements.

We examine the results of the configurations in the following order: dynamic analysis, the SMT-based refutation engines, and the parallel setup.

\subsubsection*{Sequential configuration \DYN;\HOU}

On the Windows machine there was a noticeable boost in performance using the \DYN;\HOU configuration (compared with the baseline \HOU). The overall run time improved from 17,806 to 15,283 seconds.
The maximum 93.58$\times$ speedup enabled the invariant generation for the kernel exhibiting this speedup to finish within 6.41 seconds rather than timing out after 600 seconds.

Slowdowns become severe when a refutation engine is unable to kill any candidates, in which case sequential composition always reduces performance.
For dynamic analysis, the magnitude of deceleration is generally dominated by the time required to interpret the loop body with the longest execution time.
Indeed, the 3.19$\times$ slowdown occurred in the case of a kernel whose loop body has a large number of statements, taking invariant generation from 22.21 to 70.77 seconds.
This shows that our heuristics to exit dynamic analysis early are not a panacea.
We believe a more valuable solution would be to start dynamic analysis only if a coarse estimate of kernel execution time falls below a certain threshold.
Nevertheless, this configuration offered the most impressive return: only 8 kernels suffer a noteworthy slowdown (none of which had inherently fast associated invariant refutation performance in any case), and the majority of kernels (202) benefited from a performance boost, 42 of which are noteworthy.

The picture is radically different on the Ubuntu machine, with a significant maximum slowdown and an overall loss in performance (16,632 instead of 15,544 seconds).
Investigating the kernel for which the maximum slowdown occurs in more detail, we found that, on the Windows machine, $\HOU$ and $\DYN;\HOU$ completed in 11.31 and 27.04 seconds, while, on the Ubuntu machine, the configurations completed in 7.49 and 324.33 seconds.
The wide disparity between times cannot be attributed to variations in execution paths during dynamic analysis, because the kernel is \emph{not} control dependent on formal parameter values or thread identifiers.
Moreover, recording the dynamic statement count, we verified that the interpreter performs the same work  on both machines---the counts matched (110,907 statements).
The slowdown is therefore a consequence of statement execution time, and ultimately due to the Common Language Runtime implementation; reaching the same conclusion as in our throughput experiment.
In spite of this handicap, $\DYN;\HOU$ still offered the best speedup, the second most speedups (184; after the $\DYN \| \SBASE \| \HOU$ configuration), and the most noteworthy speedups.

Our observations of dramatically different results between platforms
emphasizes the importance of accounting for measurement bias when
conducting experimental studies~\cite{MeasurementBias}, which we have attempted to
do by reporting experiments on machines with different operating
systems and runtime implementations.

\subsubsection*{Sequential configurations \{\SBASE, \SSTEP, \LU{1}\};\HOU}

\SBASE;\HOU is the only sequential configuration that offered an average cross-platform performance boost (17,428 instead of 17,806 seconds on the Windows machine, and 15,333 instead of 15,544 seconds on the Ubuntu machine).  This matches our expectations given the high throughput observed for \SBASE in the experiments of Section~\ref{sec:throughput}.

\SSTEP;\HOU offered very little, consuming the most time on both machines, amassing the fewest speedups on both machines, and creating the worst slowdown on the Windows machine.  Given the low throughput associated with \SSTEP in the experiments of Section~\ref{sec:throughput}, this is hardly surprising.

\LU{1};\HOU is similar to \SBASE;\HOU except that the former resulted in a few extra noteworthy speedups and a better maximum speedup on both machines, while the latter resulted in more speedups in total and consumed less time overall.
Given that the throughput of \SBASE was approximately double that of \LU{1}, these results suggest that \LU{1} kills more valuable candidates, leaving Houdini with less work to do.
The work that \SBASE undertakes is still worthwhile (every discovered unprovable candidate is valuable), but it leaves the more challenging unprovable candidates to Houdini.

Finally, we note that all of the SMT-based configurations negatively impacted the majority of fast kernels on both the Windows and Ubuntu machine.
Hence, these refutation engines should only be enabled when GPUVerify is likely to struggle with a kernel and its candidates.

\subsubsection*{Parallel configuration $\DYN\|\SBASE\|\HOU$}

In comparison with the other configurations, there was a marked improvement in average performance for the $\DYN\|\SBASE\|\HOU$ configuration, with a 1.27$\times$ speedup on the Windows machine and a 1.11$\times$ speedup on the Ubuntu machine.
This met our expectation that execution of Houdini in parallel with the most powerful refutation engines is superior to Houdini in isolation.

Some of the other results, however, appear counter-intuitive.
We might expect parallelization to completely eliminate the
possibility of multiple strategies slowing down invariant generation:
modulo experimental error and the modest overheads of parallelism, it
might seem that the performance of regular Houdini should be an upper
bound on parallel performance.  However, we find that worst-case slowdowns are reduced (from $3.19\times$ to $3.15\times$ on Windows, and from $43.31\times$ to $3.17\times$ on Linux), but not eliminated.
The reason is that Houdini is not impervious to the other refutation engines: \emph{how} the fixpoint is reached is influenced by the order in which refutations are discovered, and alternate orderings create variations in processing time.

\section{Related work}
\label{sec:related}

In the same vein as the GPUVerify project, several other methods for testing and verifying properties of GPU kernels have been proposed.  These include approaches based on dynamic analysis~\cite{CUDAMEMCHECK,DBLP:conf/ppopp/ZhengRQA11,TestAmplification}, verification via SMT solving~\cite{DBLP:conf/sigsoft/LiG10,DBLP:conf/sac/PereiraAMSCCSF16,HotPar}, symbolic execution~\cite{DBLP:conf/ppopp/LiLSGGR12,DBLP:journals/tse/CollingbourneCK14} and program logic~\cite{BYTECODE13,HoareSIMT}.  Among these approaches, GPUVerify is the only technique that uses candidate-based invariant generation as part of its analysis method.

Invariant generation has been a long-standing challenge in computer science that has received a lot of attention from researchers, e.g.~\cite{DBLP:conf/cav/GuptaR09,DBLP:conf/fase/KovacsV09,AI77,Houdini01,DBLP:conf/date/Eijk98,DBLP:conf/cav/JeannetM09,AbstractHoudini,DBLP:conf/cav/McMillan06,Abduction13,DBLP:conf/popl/JeannetSS14,DBLP:journals/tse/ErnstCGN01} (by no means an exhaustive list).  We discuss the work most closely related to our study.

\subsection{Candidate-based invariant generation}
Houdini was proposed as an annotation assistant for the ESC/Java tool~\cite{Houdini01}, and is formally presented in~\cite{DBLP:journals/ipl/FlanaganJL01}.  The method is analogous to an invariant strengthening technique for circuit equivalence checking~\cite{DBLP:conf/date/Eijk98}; we believe the methods were discovered independently.  Houdini can be viewed as a special instance of predicate abstraction~\cite{DBLP:conf/cav/GrafS97}, restricted to conjunctions of predicates.  This restriction is what makes the runtime of Houdini predictable, involving a worst case number of solver calls proportional to the number of candidates.  The restriction also makes it impossible to synthesize disjunctive invariants over predicates using Houdini.
A recent compelling application of Houdini is in the Corral reachability checker, where Houdini is used to generate procedure summaries which in turn are used to guide the search for bugs~\cite{DBLP:conf/cav/LalQL12}.

\subsection{Abstract interpretation}
Abstract interpretation~\cite{AI77} is a general program analysis framework that can be parameterized to generate inductive invariants over a given abstract domain.  For instance, the Interproc analyzer synthesizes invariants over the abstract domain of linear inequalities, using the Apron library~\cite{DBLP:conf/cav/JeannetM09}.  Predicate abstraction is abstract interpretation over the domain of Boolean combinations of predicates~\cite{DBLP:journals/sttt/BallPR03}, and Houdini is thus a form of abstract interpretation where the domain is restricted to conjunctions of predicates.  The main disadvantages of abstract interpretation are that it is inflexible, in the sense that generation of invariants beyond a given abstract domain requires a bespoke new domain to be crafted, and that to ensure convergence to a fixpoint it is necessary to apply \emph{widening} which can be hard to control in a predictable manner.  In contrast, a Houdini-based approach can easily be  ``tweaked'' by adding new candidate generation rules on an example-driven basis, as we have demonstrated in this paper.  Convergence to a fixpoint is also predictable based on the known set of candidates.  In recent work, \emph{Abstract Houdini} has been proposed in an attempt to combine the benefits of abstract interpretation and candidate-based invariant generation~\cite{AbstractHoudini}.

\subsection{Invariant generation for affine programs}
There has been significant progress recently on invariant generation for a restricted class of programs that operate on unbounded integers and only compute affine expressions over program variables.  Under these restrictions, novel applications of Craig interpolation~\cite{DBLP:conf/cav/McMillan06}, abduction~\cite{Abduction13} and abstract acceleration~\cite{DBLP:conf/popl/JeannetSS14} have been shown to be effective in invariant synthesis. The weakness of these methods are the restrictions on input programs.  In our application domain, for example, programs operate on fixed-width bit-vectors and floating point numbers.  It is necessary to reason precisely about bit-vectors to capture arithmetic using powers-of-two, frequently encoded efficiently using shifting and masking, and we require support for uninterpreted functions to abstract floating point operators but retain their functional properties.  Furthermore, GPU kernels frequently exhibit non-linear computations.  For example, reduction operations involve loops in which a counter exponentially varies in powers of two between an upper and lower bound.  These characteristics render methods for affine programs inapplicable in our setting.

\subsection{Dynamic invariant generation}
The techniques discussed above all use static analysis to establish program invariants with certainty.  In contrast, dynamic invariant generation, pioneered by the Daikon system~\cite{DBLP:journals/tse/ErnstCGN01} employs dynamic analysis with respect to a test suite to speculate \emph{likely} invariants: facts that are found to hold invariantly during testing, with statistical evidence that the dynamic invariance of these facts appears to be non-coincidental.  This method provides no guarantee that the suggested facts are actually invariants.  A study combining the Daikon method with extended static checking for Java considered the use of dynamically generated invariants as a source of candidates for Houdini~\cite{DBLP:conf/sigsoft/NimmerE02}.

\subsection{Studies on invariant generation}
A related study on invariant generation~\cite{Polikarpova09} aimed to evaluate whether it is better to rely on manual effort, automated techniques or a combination of both in generating program invariants.
The study concludes that a combination is required: Daikon inferred $5$ times as many invariants as specified manually, but could only find approximately $60\%$ of the manually crafted invariants.
The benchmark set consisted of $25$ classes taken partially from widely used libraries and partially written by students.
The size of the benchmark set allowed the authors to investigate each inferred assertion individually; this is not feasible in our study due to the substantially larger number of benchmarks.

\section{Conclusions}
\label{sec:conclusions}

In this study we have shown that candidate-based invariant generation is valuable to GPUVerify, significantly increasing the precision of the tool and, to some extent, relieving the burden of manual loop-invariant discovery.
This success is in large part due to our strategy of incorporating new rules into GPUVerify because candidate-based invariant generation is only as good as the supply of speculated candidates.
However, our evaluation also provides a cautionary tale: rules may become redundant over time, particularly when new rules are introduced, thus a continual assessment of their use in the verification tool is worthwhile.

The wider issue with candidate-based invariant generation is that, in general, more rules mean more candidates and, ultimately, more processing time.
The refutation engines and the infrastructure that we implemented to curb processing time proved effective when comparing invariant discovery with and without these techniques. Our mechanism to choose between refutation engines and between sequential or parallel processing mainly rested on empirical evidence of throughput and complementary power.
The drawback of this, as the results indicate, is that the unselected refutation engines or processing modes could be better for \emph{specific} kernels.
As is, our setup ignores all properties of the program and of the candidate invariants.
Future work may therefore investigate machine learning techniques to fine-tune the setup.
Another avenue for future work is to investigate additional parallel refutation strategies, in addition to the strategy that we predicted to be the most promising.

\section*{Acknowledgements}

This work was supported by the EU FP7 STREP project CARP (project
number 287767), the EPSRC PSL project (EP/I006761/1), Imperial College
London's EPSRC Impact Acceleration Account, and a gift from Intel
Corporation.



\bibliographystyle{IEEEtran}
\bibliography{IEEEabrv,references}

%

\begin{IEEEbiography}[{\includegraphics[width=1in,clip,keepaspectratio]{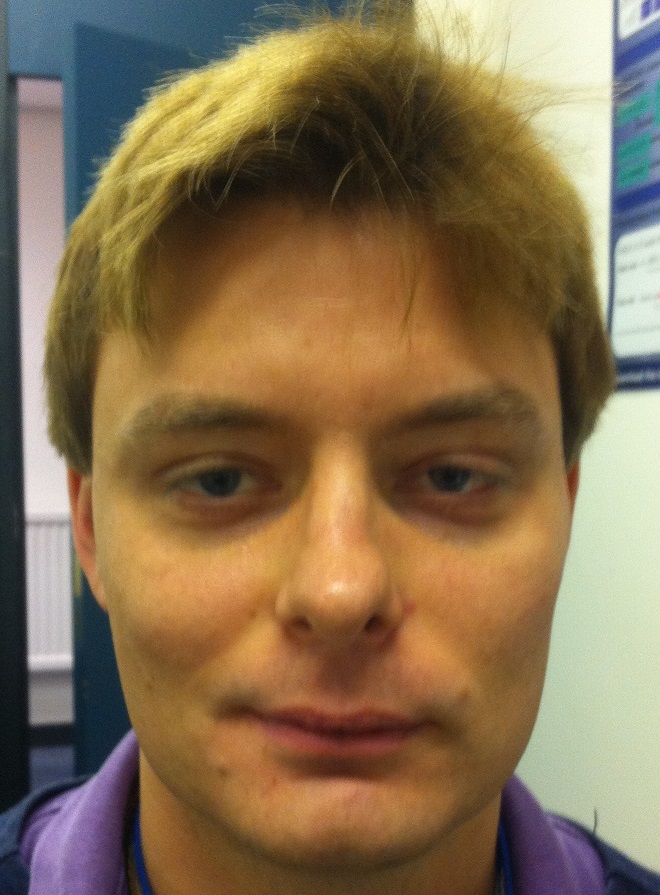}}]{Adam Betts}
holds a PhD from the University of York in worst-case execution time analysis, and contributed to this paper while a postdoctoral researcher in the Multicore Programming Group at Imperial College London, as part of the CARP EU FP7 project.
\end{IEEEbiography}
\begin{IEEEbiography}
    [{\includegraphics[width=1in,clip,keepaspectratio]{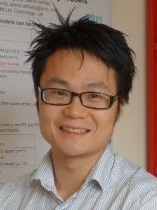}}]{Nathan Chong}
contributed to this work while he was a PhD student in the Multicore Programming Group at Imperial College London.  His PhD thesis, ``Scalable Verification Techniques for Data-Parallel Programs'', developed a number of approaches to reasoning about GPU kernels, with associated publications at leading conferences, including OOPSLA and POPL, and to him winning the UK ICT Pioneers competition in 2014.  He is currently a researcher at ARM, where he works on hardware and software at many levels of the system stack.
\end{IEEEbiography}
\begin{IEEEbiography}
    [{\includegraphics[width=1in,clip,keepaspectratio]{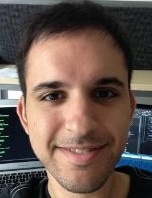}}]{Pantazis Deligiannis}
was a PhD student at Imperial College London, writing a thesis on ``Scalable Techniques for Analysing and
Testing Asynchronous Software Systems''.  His PhD work included the development of the P\# language and toolchain for building and reasoning about asynchronous systems on top of the .NET framework, leading to papers at top conferences such as PLDI and FAST, and to him reaching the final of the UK ICT Pioneers competition in 2015.  During his PhD he also contributed to the GPUVerify project.  Pantazis is currently a Research Software Development Engineer at Microsoft Research in Cambridge.
\end{IEEEbiography}
\begin{IEEEbiography}
    [{\includegraphics[width=1in,clip,keepaspectratio]{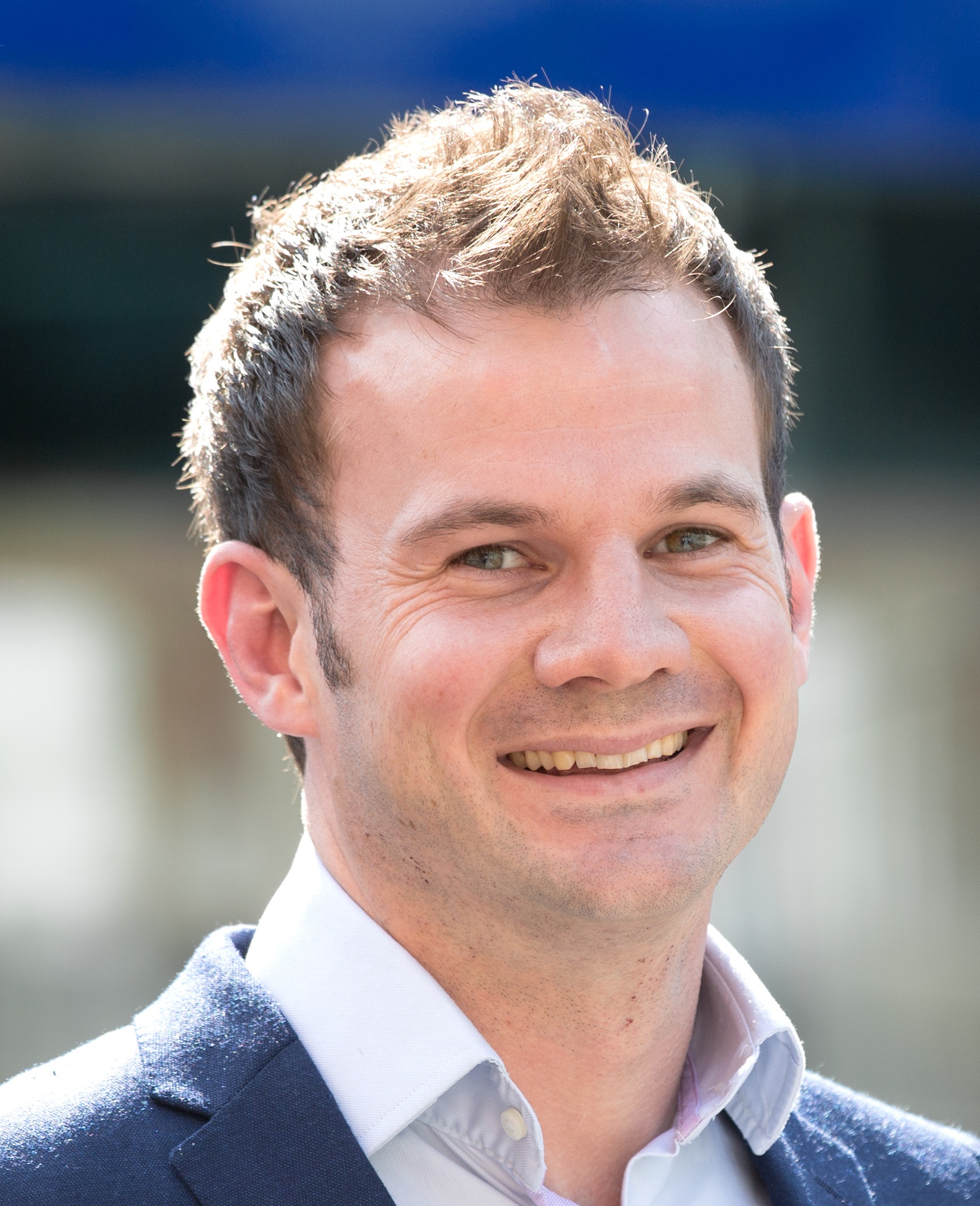}}]{Alastair F. Donaldson}
is a Senior Lecturer at Imperial College London, where he leads the Multicore Programming Group and holds an EPSRC Early Career Fellowship.  He has published more than 70 peer-reviewed articles in the fields of Programming Languages, Verification, and Testing, and his contributions to many-core programming led to him receiving the 2017 British Computer Society Roger Needham Award.
\end{IEEEbiography}

\newpage
\begin{IEEEbiography}[{\includegraphics[width=1in,clip,keepaspectratio]{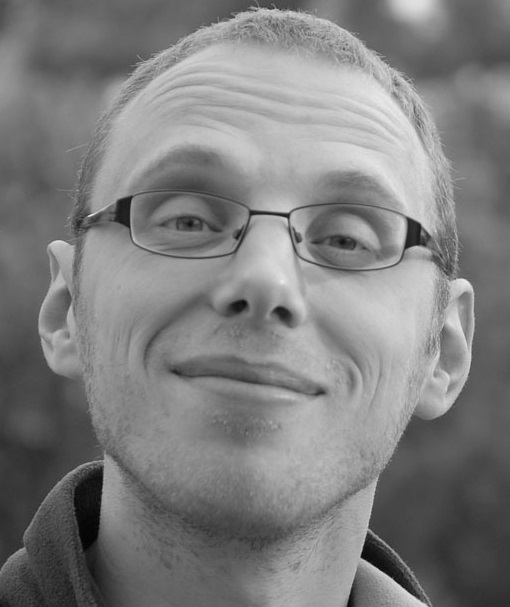}}]{Jeroen Ketema}
holds a PhD in theoretical computer science from the Vrije Universiteit, Amsterdam and contributed to this work while he was a postdoctoral researcher in the Multicore Programming Group at Imperial College London, working on the CARP EU FP7 project. He has published in the areas of Verification and the Semantics of Programming Languages. Jeroen is currently a Research Fellow at the Netherlands Organization for Applied Scientific Research (TNO).
\end{IEEEbiography}





\clearpage

\appendices{}
\section{Invariant generation rules in GPUVerify}
\label{app:rules}

The invariant generation rules broadly fall into the following categories:
\begin{itemize}
  \item Patterns over accesses.
  These summarize reads and writes that are issued when executing a loop.
  Examples include strides or ranges of accesses and can include one, two, or three-dimensional patterns.

  \item Patterns over loop guards.
  These summarize ranges and values that variables used in loop guards may assume.

  \item Variables that are always powers of two.
  These are useful in kernels that perform tree reductions or prefix sums~\cite{Blelloch90}, where the variable is used as an offset for issuing reads and writes and to disable threads that no longer take part in a calculation.
  Sometimes it is necessary to discover relationships between pairs of power-of-two variables, e.g.\ one power-of-two variable doubling on every loop iteration, and another halving on every loop iteration, so that their product is invariant.

  \item No accesses can be in-flight at loop heads.
  These make judgments concerning reads or writes issued during the execution of a loop.


  \item Uniformity of variables across threads.
  These are needed to verify absence of barrier divergence.
  They establish that threads follow the same flow of control, or that variables have the same value at a loop head.
\end{itemize}
We now give a brief description of each rule, in some cases presenting a motivating kernel fragment.
In the following, $\mathsf{write}(\verb!out!)$ denotes the set of writes that have been issued to array \verb!out! and that may still be in-flight; $\mathsf{read}(\verb!out!)$ is similar.  In addition, $C$ denotes a constant value and $e$ an expression.
Several of the rules refer to threads being \emph{enabled} and \emph{disabled}.  The notion of enabledness is part of the predicated semantics that GPUVerify uses when translating a GPU kernel into Boogie~\cite{TOPLAS15,GPUVerify13}.

\subsection{r0. accessBreak}
Given an access pattern involving thread \emph{components} (thread or block identifiers), this rule attempts to break the access pattern into its component forms using rewriting.
The rule is based on the intuition that relating an access pattern of a thread to its components is useful because each thread must always have at least one component that is unique to that thread when compared to another thread.

\subsection{r1. accessedOffsetsSatisfyPredicates}
This rule identifies stride patterns and strength reduction loops.
For the loop
\begin{lstlisting}
for (int i = 0; i < 4; i++) {
  out[i*blockDim.x+threadIdx.x] = ...
}
\end{lstlisting}
the rule generates:
\begin{multline*}
\forall w \in \mathsf{write}(\code{out}). \\ (w \; \% \; \code{blockDim.x}) = \code{threadIdx.x} \, .
\end{multline*}

\subsection{r2. accessLowerBoundBlock and r3. accessUpperBoundBlock}
These rules identify whether a block of threads is assigned a contiguous range of an array. The candidates generated are lower and upper bounds that restrict the range of accesses.
For the loop
\begin{lstlisting}
for (int i = 0; i < 4; i++) {
  out[$C$*blockDim.x+i] = ...
}
\end{lstlisting}
the rules generate:
\[\forall w \in \mathsf{write}(\code{out}) . \; C\cdot\code{blockIdx.x} \le w\]
and
\[\forall w \in \mathsf{write}(\code{out}). \; w < C\cdot(\code{blockIdx.x}+1) \le w \, .\]

\subsection{r4. accessOnlyIfEnabledInEnclosingScopes}
The rule specifies that accesses can only have been issued when a thread is enabled in all enclosing scopes. For
\begin{lstlisting}
if (x < k) {
  if (y < l) {
    for (...) {
      out[...] = ...;
    }
  }
}
\end{lstlisting}
the rule generates:
\[\mathsf{write}(\code{out}) \neq \emptyset \Rightarrow \code{x} < \code{k} \; \land \; \code{y} < \code{l} \, .\]

\subsection{r5. conditionsImplyingEnabledness}
This rule specifies that a thread is only enabled when it is enabled in all enclosing scopes.

\subsection{r6. disabledMaintainsInstrumentation}

This rule specifies that if a thread is disabled during the execution of a region of code, then the accesses that are being tracked for the thread cannot change as a result of the execution of this region.

\subsection{r7. guardMinusInitialIsUniform}
This rule specifies that loops containing barriers must have \emph{uniform} conditions, meaning that the loop condition evaluates to the same value across all threads.  This is necessary to avoid barrier divergence.
A common case that makes this non-trivial to prove is where the loop counter is initialized to a non-uniform value, typically a thread identifier, and is thereafter incremented by a uniform value, often a thread block dimension.  In this case, knowing that the loop counter minus its initial value is uniform may suffice to prove that the Boolean value of the loop guard is uniform.
For the loop
\begin{lstlisting}
for (int i = threadIdx.x;
       i < N; i += blockDim.x) {
  __syncthreads();
}
\end{lstlisting}
the rule generates:
\[\mathsf{uniform}(\code{i} - \code{threadIdx.x})\,,\]
which specifies that the expression $\code{i} - \code{threadIdx.x}$ is uniform.

\subsection{r8. guardNonNeg}
This rule specifies that every guard variable is non-negative.
For the loop
\begin{lstlisting}
for (int i = C; i > 0; i--) {
  ...
}
\end{lstlisting}
the rule generates $0 \le \code{i}$.

\subsection{r9. loopBound}
This rule specifies that the initial value of a guard variable bounds the range of the guard.
For the loop
\begin{lstlisting}
for (int i = $e$; ...) {
  ...
}
\end{lstlisting}
the rule generates $e \le \code{i}$ and $\code{i} \le e$.

\subsection{r10. loopCounterIsStrided}
This rule is analogous to accessedOffsetsSatisfyPredicates but for guard variables.

\subsection{r11. loopPredicateEquality}
This rule is concerned with barrier divergence and uniformity of control flow, and is intimately related to the predicated execution semantics that GPUVerify employs when translating a kernel into Boogie form.

\subsection{r12. noread and r13. nowrite}
These rules identify that no accesses can be in-flight at the loop head when a barrier appears in the loop.
For
\begin{lstlisting}
for (...) {
  __syncthreads();
}
\end{lstlisting}
\newpage
\noindent
the rules generate:
\[\mathsf{read}(\code{out}) = \emptyset\]
and
\[\mathsf{write}(\code{out}) = \emptyset \, .\]

\subsection{r14. pow2 and r15. pow2NotZero}
These rules identify variables that only assume power-of-two values.
The rule is based on the intuition that power-of-two values are often used in bit masks, tree reductions, and prefix sums~\cite{Blelloch90}.
For the loop
\begin{lstlisting}
for (int x = N; x > 0; x >>= 1) {
  ...
}
\end{lstlisting}
the rule pow2 will generate
\[
\code{x} = 0 \; \lor \; \code{x} = 1 \; \lor \; \cdots \; \lor \; \code{x} = 2^{31} \, ,
\]
while pow2NotZero will generate the same disjunction but excluding the $\code{x} = 0$ case.

\subsection{r16. predicatedEquality}
This rule is concerned with barrier divergence and uniformity of variables, and is intimately related to the predicated execution semantics that GPUVerify employs when translating a kernel into Boogie form.

\subsection{r17. relationalPow2}
This rule identifies possible pairs of power-of-two variables where one is being incremented and the other is being decremented.
It specifies a lock-step relation between the variables.

\subsection{r18. sameWarpNoaccess}

This rule concerns threads that are in the same \emph{warp}---a unit
of typically 32 threads that, on NVIDIA architectures, execute in lock
step.  GPUVerify provides optional support for warps~\cite{EthelNFM}.  If this support
is enabled, the rule speculates that if the two threads under
consideration are in the same warp, then neither thread has any
pending in-flight accesses (lock-step execution guarantees
that the threads synchronize with one another after each instruction).

\end{document}